\documentclass{aa}

\usepackage[utf8]{inputenc} 
\usepackage[varg]{txfonts}
\usepackage{amssymb}
\usepackage{epsfig}
\usepackage{graphicx} 
\usepackage{amsmath}
\usepackage{color}
\usepackage{natbib}
\usepackage{hyperref}
\usepackage{gensymb}

\newcommand{\bmath}[1]{\boldsymbol{#1}}

\def\bc{\begin{center}}
\def\ec{\end{center}}
\def\beq{\begin{eqnarray}}
\def\eeq{\end{eqnarray}}

\def\d{d}
 
\def\rg{r_{\rm S}} 
\def\Dop{\delta}
\def\source{\object{SAX J1808.4$-$3658}}
\def\req{R_{\mathrm{eq}}}

\bibpunct{(}{)}{;}{a}{}{,} 

\DeclareUnicodeCharacter{00A0}{ }

\newcommand{\be}{\begin{equation}}
\newcommand{\ee}{\end{equation}}
\newcommand{\bea}{\begin{align}\begin{split}}
\newcommand{\eea}{\end{split}\end{align}}

\newcommand{\msun}{{M}_{\sun}}

\makeatletter
\def\fvec#1{\underline{\sbox\tw@{$#1$}\dp\tw@\z@\box\tw@}}
\makeatother

\begin{document}

\title{Bayesian parameter constraints for neutron star masses and radii using X-ray timing observations of accretion-powered millisecond pulsars}

\titlerunning{Bayesian pulse profile modelling of accreting millisecond pulsars}

\author{T.~Salmi\inst{1}
\and J.~N\"attil\"a\inst{2}
\and  J.~Poutanen\inst{1,2,3}}

\institute{Tuorla Observatory, Department of Physics and Astronomy, FI-20014 University of Turku, Finland 
 \and Nordita, KTH Royal Institute of Technology and Stockholm University, Roslagstullsbacken 23, SE-10691 Stockholm, Sweden
   \and Space Research Institute of the Russian Academy of Sciences, Profsoyuznaya str. 84/32, 117997 Moscow, Russia 
}

\date{Received 3 May 2018 / Accepted 25 August 2018}

\abstract{
We present a Bayesian method to constrain the masses and radii of neutron stars (NSs) using the information encoded in the X-ray pulse profiles of accreting millisecond pulsars.
We model the shape of the pulses using ``oblate Schwarzschild'' approximation, which takes into account the deformed shape of the star together with the special and general relativistic corrections to the photon trajectories and angles.
The spectrum of the radiation is obtained from an empirical model of Comptonization in a hot slab in which a fraction of seed black-body photons is scattered into a power-law component. 
By using an affine-invariant Markov chain Monte Carlo ensemble sampling method, we obtain posterior probability distributions for the different model parameters, especially for the mass and the radius.
To test the robustness of our method, we first analysed self-generated synthetic data with known model parameters.
Similar analysis was then applied for the observations of SAX J1808.4$-$3658 by the {\it Rossi} X-ray Timing Explorer (RXTE). 
The results show that our method can reproduce the model parameters of the synthetic data, and that accurate constraints for the radius can be obtained using the RXTE pulse profile observations if the mass is a priori known.
For a mass in the range $1.5{-}1.8~\msun$, the radius of the NS in SAX J1808.4$-$3658 is constrained between 9 and 13~km.
If the mass is accurately known, the radius can be determined with an accuracy of 5\% (68 \% credibility). 
For example, for the mass of $1.7~\msun$ the equatorial radius is $\req = 11.9^{+0.5}_{-0.4}$ km. 
Finally, we show that further improvements can be obtained when the X-ray polarization data from the Imaging X-ray Polarimeter Explorer will become available.
}


\maketitle
 
\section{Introduction}\label{sec:intro}
 
Neutron stars (NSs) are the densest directly observable objects in the Universe. 
The matter inside a NS is at supranuclear densities. 
The pressure-density-temperature relation of such dense matter is described by the equation of state. 
There is one-to-one mapping relating the equation of state to the mass--radius dependence of the NS \citep[see, for example,][]{lindblom1992,Lattimer12ARNPS}. 
Thus, measurements of the NS masses and radii from observations can in principle help us to constrain microphysical properties of high-density matter.  
One way to obtain information on these parameters is to study pulse profiles produced by hotspots on the surface of rapidly spinning NSs \citep{2016RvMP...88b1001W}, in particular, in accreting millisecond pulsars (AMPs).

In accreting millisecond pulsars, the matter from a low-mass companion star accretes onto the magnetic poles of a rapidly rotating NS, forming two hotspots on the stellar surface. 
Because the hotspots are misplaced from the rotational axis of the NS, the X-ray radiation observed from these spots pulsates coherently at the spinning frequency of the NS. 
The pulse profiles carry information about the mass and radius of a NS because the light bending and thus pulse shape depends strongly on the compactness of the star \citep{P08}.
However, many other physical parameters and observing angles affect both the light curves and the spectra of these sources.

To model the pulse profiles from the hotspots of NSs, the necessary formalism was first developed by \citet{PFC83} and \citet{Page95}.
For rapid rotation the formalism was extended by \citet{ML98}, \citet{WM01}, and \citet{PG03}.  
The transformation of polarization was studied by \citet{VP04}. 
A simple approximation for light bending was given in \citet{b02}. 
Other useful analytical results were obtained by \citet{PB06}. 
How deviations from the Schwarzschild  metric affect the profiles have been studied by \citet{BRR00} (using the Kerr metric), who corrected previous errors by \citet{CS89} and showed that the effects are rather small. 
The effects of oblateness of the star due to rapid rotation were studied by \citet{MLC07} using Schwarzschild metric and by \citet{CMLC07} using a numerically-generated metric for rapidly rotating NSs in general relativity. 
A ray-tracing algorithm using Hartle-Thorne metric was examined by \citet{BPOJ12}.
In addition, \citet{NP18} showed that the special relativistic rotational effects emerge directly from a general relativistic treatment of a rotating compact object.

In order to infer the stellar mass and radius, with help of the forthcoming observations, the pulse profiles were investigated by \citet{POC14} using approximate relations between the modelled and observed parameters. 
Previously, mass and radius constraints using pulse profile modelling for thermonuclear burst oscillations and Bayesian analysis were studied by \citet{lomiller13} and \citet{ML15}. 
Mass-radius constraints were also studied by \citet{Stevens2016} using evolutionary optimization algorithm. 
They as well considered the thermonuclear burst oscillations and used the synthetic data, however, keeping spot size and temperature fixed. 
  
In this paper, our aim is to get new information of mass-radius relation using the energy-resolved pulse profiles of accretion-powered millisecond pulsations, for which we have plenty of already existing data \citep[see e.g.][]{GDB02,PG03,GP05,LMC08,ML11}. 
On the other hand, we also want to show how accurately NS parameters could be constrained with help of the upcoming polarimeters and X-ray missions like the Imaging X-ray Polarimeter Explorer (IXPE) \citep{IXPE} and the enhanced X-ray Timing and Polarimetry mission (eXTP) \citep{EXTP}.
We do the full Bayesian analysis with both synthetic data and observations of \source, and fit the data simultaneously in phase and energy dimensions. 
Like {\citet{ML15}}, we sample the parameter space for synthetic data using a Markov Chain Monte Carlo (MCMC) method.
However, we have more free parameters, and instead of maximizing, we also marginalize the likelihoods over the distance and spot temperature. 
In particular, we now have the angular dependence of the emitted radiation (beaming) as a free parameter. 
 
The remainder of this paper is structured as follows. 
In Sect. \ref{sec:methods}, we describe the methods we have used to fit the pulse profiles and spectra of accreting millisecond pulsars. 
In Sect. \ref{sec:data} we present our synthetic data as well as the original data for \source\ that we have used in our analysis. 
The results of the modelling are described in Sect. \ref{sec:results}. 
We discuss the results in Sect. \ref{sec:discussion} and summarize in Sect. \ref{sec:summary}.  

\section{Methods}\label{sec:methods}

\subsection{Pulse shape modelling}\label{sec:pulses}
 
Our pulse shape modelling is based on the model introduced in {\citet{PB06}}, which takes the special and general relativistic effects into account by using the `Schwarzschild-Doppler' approximation. 
The effects of general relativity (gravitational light bending) are modelled as though the star is not rotating describing the exterior metric with a Schwarzschild solution.
On the other hand, rotational effects have been approximated by using special relativistic Doppler transformations and angle aberrations as though the star is a rotating object with no gravitational field.

\begin{figure}
\resizebox{\hsize}{!}{\includegraphics{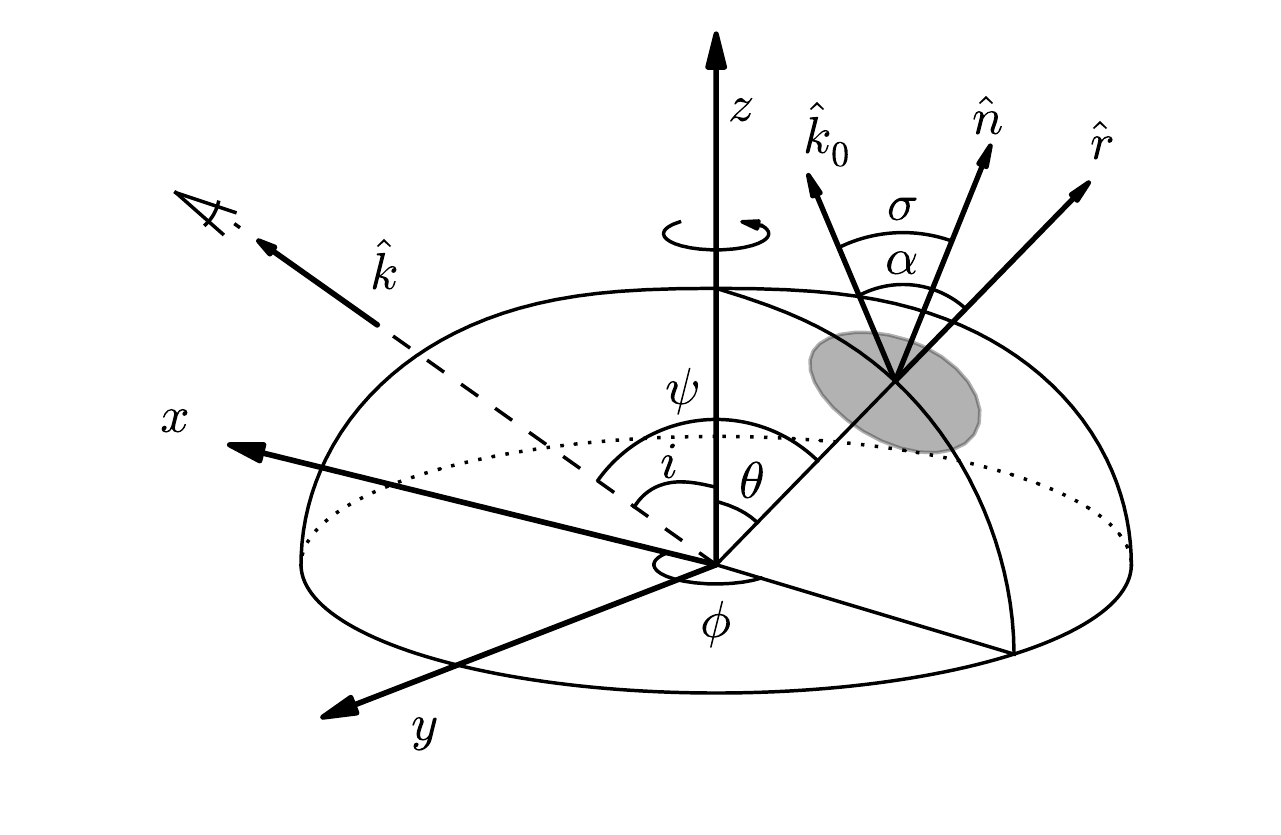}}
\caption{Geometry of the problem.
The Cartesian ($x, y, z$) and spherical (with $\theta$ and $\phi$) coordinate systems are shown.
The star has an oblate shape, because it is rotating rapidly around the $z$-axis.
Therefore, the radial vector $\hat{r}$ and the surface normal $\hat{n}$ differ from each other.
The photon from the hot spot is initially emitted towards $\hat{k}_{0}$ but due to the light bending it reaches the observer in direction of $\hat{k}$.
} 
\label{fig:geom2}
\end{figure}

In Schwarzschild geometry we know the exact relation between the photon emission angle $\alpha$ and the deflection angle $\psi$. 
The angle $\alpha$ is measured between the radial direction $\bmath{r}$ and the initial direction of photon $\bmath{k_{0}}$ (see Fig. \ref{fig:geom2}), and the angle $\psi$ between $\bmath{r}$ and observer $\bmath{k}$ (see Fig. \ref{fig:geom2}). 
When $\alpha < \pi/2$, it is given by \citep[e.g.][]{mtw73,PFC83,b02}
\be \label{eq:bend2}
  \psi_{\mathrm{p}}(R,\alpha)=\int_R^{\infty} \frac{dr}{r^2} \left[ \frac{1}{b^2} -
       \frac{1}{r^2}\left( 1- \frac{\rg}{r}\right)\right]^{-1/2} ,
\ee
where $b$ is the impact parameter,
\be \label{eq:impact2}
  b=\frac{R}{\sqrt{1-u}} \sin\alpha,
\ee
$u\equiv\rg/R$, $\rg=2GM/c^2$ is the Schwarzschild radius, $G$ is the gravitational constant, $c$ is the speed of light, $R$ is the radius of the star at the point which emits the photon, and $M$ is the mass of the NS.  
A numerical method to evaluate integral (\ref{eq:bend2}) is presented in Appendix \ref{sec:appendix}. 
For an oblate star, there is a possibility to have photon trajectories with decreasing radial coordinate (i.e. photons emitted towards the star) corresponding to $\alpha > \pi/2$. 
In this case, the relation between $\alpha$ and $\psi$ takes the form
\be 
\psi(R,\alpha)=2\psi_{\mathrm{\max}}-\psi_{\mathrm{p}}(R,\pi-\alpha),
\ee 
where $\psi_{\max} = \psi_{\mathrm{p}}(p,\alpha=\pi/2)$ and $p$ is the distance of closest approach, given by
\be
p = -\frac{2}{\sqrt{3}}b\cos\left([\arccos{(3\sqrt{3}\rg/2b)}+2\pi]/3\right).
\ee

The original model of \citet{PB06} has been expanded to take into account the geometrical effects of the oblate shape of the NS as described by \citet{MLC07}. 
The oblate shape of the star is obtained by using a function from \citet{AGM14}, which gives the radius $R$ of the star as a function of co-latitude. 
The geometry of the system with the star, hot spot, and observer in the direction $\bmath{k}$ is presented in Fig.~\ref{fig:geom2}. 
Here $\theta$ is the co-latitude of the spot, $i$ is the observer inclination, and $\phi$ is the spot phase angle. 
Due to the oblateness, the local surface normal $\bmath{n}$ and radius vector $\bmath{r}$ are not always pointing to the same direction. 
Therefore, to check the visibility condition, we need to consider the emission angle $\sigma$ relative to $\bmath{n}$, instead of $\alpha$. 
Only photons with emission angle $\sigma\leq\pi/2$ can reach the observer.
However, the emitted photon might not be visible, even though $\cos \sigma>0$, if the photon hits the NS surface at any later phase of its trajectory. 
We tested this requirement empirically by studying different photons emitted almost parallel to the surface from various emission locations. 
Photons were propagated from the surface and 
their radial location was tested to see if the trajectory intersected with the oblate surface. 
These tests show that (at least for all the cases we considered) a photon initially satisfying the requirement $\cos \sigma >0$ will always reach the distant observer.

Hereafter, all the primed quantities are measured in the co-rotating frame of the spot.
The infinitesimal spot area measured in the frame co-moving with the spot may be calculated as \citep{MLC07} 
\be \label{eq:surf_element}
\d S'(\theta) = \gamma R^{2}(\theta) [1+f^{2}(\theta)]^{1/2} \sin \theta \d \theta \d \phi,
\ee
where the Lorentz factor $\gamma = 1/\sqrt{1-\beta^{2}}$,
$\beta$ is the spot velocity in units of $c$,
\be
f(\theta)=\frac{1+z}{R}\frac{\d R}{\d \theta},
\ee
and $1+z = 1/\sqrt{1-u}$ is the gravitational redshift. 
The factor $[1+f^{2}(\theta)]^{1/2}$ takes into account the oblateness of the spot surface. 
The Lorentz factor $\gamma$ originates from considering a differential area element in rotating coordinates.
It results from the fact, that the surface element $\d S'$ here is defined based on simultaneous measurements of co-moving observers, instead of using simultaneous measurements of local static observers \citep{NP18, lo_err18}.

Following derivation presented by \citet{PB06}, we can get the observed spectral flux 
\be \label{eq:fluxspot2}
\d F_{E}=(1-u)^{1/2} \Dop^{4} I'_{E'}(\sigma') \cos\sigma
\frac{\d \cos\alpha}{\d\cos\psi}
 \frac{\d S'}{D^2} ,
\ee
where $D$ is the distance to the star, $I'_{E'}$ is the spectral intensity and $\sigma'$ is the emission angle relative to the normal as measured in the co-moving frame of the spot,
\be \label{eq:dop2}
\Dop=\frac{1}{\gamma(1-\beta\cos\xi)} 
\ee
is a Doppler factor, and $\xi$ is the angle between the spot velocity and initial direction of the photon. 
We note that the only difference from Equation (A15) given in \citet{PB06} is the change of $\alpha'$ to $\sigma'$ in the argument of the intensity and in $\cos\sigma$, which accounts for the projection of the spot area on the observer sky.

The flux in expression \eqref{eq:fluxspot2} depends on the pulsar phase $\phi$ (because e.g. the Doppler factor $\delta$ depends on it). 
This flux actually corresponds to an observed phase $\phi_{\rm obs}$, which is different from $\phi$ due to light travel delays. 
The time delay is caused by different travel times of emitted photons to the observer, depending on the position of the emitting point.
A photon following the trajectory with an impact parameter $b$ (and $\alpha < \pi/2$) is lagging a photon originating at the same radius $R$ with impact parameter $b=0$ by  \citep{PFC83} 
 \be \label{eq:timedelay}
c\Delta t_{\mathrm{p}}(R,\alpha) =
\int_R^{\infty} \frac{\d r}{1- \rg/r}
\left\{ \left[ 1-  \frac{b^2}{r^2}  \left( 1- \frac{\rg}{r} \right)
\right] ^{-1/2}  -1 \right\} . 
\ee
Because we need to account for the stellar oblateness, we have to compute all the delays relative to the photons emitted at some reference radius $r_{\mathrm{ref}}$ (which can be chosen arbitrarily as long as it exceeds $\rg$) and  $b=0$ (i.e. $\alpha=0$). 
Photons with $b=0$ emitted at radius $R$ will be lagging photon emitted at $r_{\mathrm{ref}}$ by 
\be
c\delta t(R,r_{\mathrm{ref}}) = r_{\mathrm{ref}}-R+\rg\ln\left(\frac{r_{\mathrm{ref}}-\rg}{R-\rg}\right) . 
\ee
Thus the total delay of photons emitted at $R$ and angle $\alpha$ (with corresponding impact parameter given by Eq.~(\ref{eq:impact2})) relative to the photons emitted at $r_{\mathrm{ref}}$ with $b=0$ is 
\be \label{eq:delay2}
\Delta t(R,\alpha)=  \Delta t_{\mathrm{p}}(R,\alpha) + \delta t(R,r_{\mathrm{ref}}) .
\ee 
In our model, we use the equatorial radius as a reference radius ($r_{\mathrm{ref}} = \req$).
A numerical method to evaluate integral (\ref{eq:timedelay}) is presented in Appendix \ref{sec:appendix}.

In the case of $\alpha > \pi/2$, the corresponding delay can be calculated using the symmetry around distance $p$ of the closest approach as
\beq\label{eq:delay22}
\Delta t(R,\alpha) &=& 2\Delta t(p,\pi/2)-\Delta t(R,\pi-\alpha) \nonumber \\
&=& 2\Delta t_{\mathrm{p}}(p,\pi/2)-\Delta t_{\mathrm{p}}(R,\pi-\alpha)  \nonumber \\ 
&+& 2\delta t(p,r_{\mathrm{ref}}) - \delta t(R,r_{\mathrm{ref}}).
\eeq

Otherwise, the computation of the light curve follows \citet{PB06} with corrections for oblateness as described in \citet{MLC07}. 
One notable difference is, however, that we are considering finite-sized spots which may be also large in size.
In order to compute the flux from a spot with finite size, we need to integrate over the spot surface, in other words we integrate Eq.~\eqref{eq:fluxspot2} with $dS'$ given by Eq.~\eqref{eq:surf_element} over the solid angle.
It is done by splitting the spot into a number of small sub-spots and computing the fluxes to each sub-spot separately. 
For a circular spot around the magnetic pole, instead of using $\theta$ and $\phi$ are variables, it is  easier
to integrate over the solid angle using Gaussian quadrature in the cosine of the magnetic co-latitude (i.e. the angle measured from the magnetic pole) and using trapezoidal rule for integration over the corresponding azimuth. 
For each phase bin and sub-spot we need to evaluate light bending and time delays separately, because each sub-spot is located at a different radius due to the oblateness.

We have also made some physical assumptions to simplify the model.
We assume a spot which has a constant angular radius $\rho$, which is the angle between the centre and the edge of the spot measured from the centre of the star. 
In this paper we also model the pulse profiles using only one hot spot \citep[see e.g.][]{IP09}. 
In addition, we ignore the radiative transfer effects related to photon propagation through the accretion stream.  
See Sect. \ref{sec:discussion} for discussion how these assumptions may affect the results.

\subsection{Spectrum of the radiation}
\label{sec:spectra}

Next we present how the energy and angular spectra of radiation are modelled, or how the intensity $I'_{E'}(\sigma')$ depends on energy $E'$ and emission angle $\sigma'$. 
The  material accreted onto a NS is abruptly decelerated close to the surface releasing its kinetic energy and forming a hotspot. 
The energy spectrum of AMPs above 3 keV can be well represented as a sum of the black-body-like component at lower energies and a power-law-like component extending to $\sim$100 keV and likely produced by thermal Comptonization \citep[e.g.][]{PG03,GP05,2008AIPC.1054..157F}. 
We can associate the black-body component with the emission from the heated NS surface and the Comptonization components with the accretion shock or very surface layers heated to 50--100 keV by bombarding particles. 
The seed photons for Comptonization are likely coming from the cooler layers immediately under the shocked plasma. 
In this work we thus use a two-component model for the spectrum.

For each point inside the spot, we represent the intensity as a sum of the black-body ($I'_{\mathrm{bb}}$) and Comptonization ($I'_{\mathrm{comp}}$) parts.  
Following \citet{PG03}, we assume that the intensities of both black-body and Comptonization components can be expressed as products of functions that depend either only on energy $E'$ or only on emission angle $\sigma'$. 
For the black-body component we use the Planck law for the specific intensity, in other words we assume the angular pattern to be isotropic, which means that $I'_{\mathrm{bb}}(\sigma')= I'_{\mathrm{bb}}$

\subsubsection{Radiation from hot spots}

In this paper we approximate the spectrum with Comptonization model \textsc{simpl} introduced by \citet{Steiner2009} and implemented into \textsc{xspec} \citep{Arn96}. 
This model is based on the solution of the Kompaneets equation describing Compton scattering in non-relativistic plasmas \citep{ST80}.
In this model a fraction of photons of a seed black-body spectrum is Compton scattered into a power-law-like component with photon spectral index $\Gamma$:
\be\label{eq:gammaphot}
 I'_{E'} \propto E'^{-(\Gamma-1)}.
 \ee
The parameters of the \textsc{simpl} model are $\Gamma$ and the scattered fraction $X_{\mathrm{sc}}$ of black-body photons.  
In addition, we take  into account only the up-scattered photons.
The model \textsc{simpl-2}, designed for non-relativistic thermal Comptonization, would take into account also the down-scattered photons.
However, since the difference between the models is small, we use the more simple model \textsc{simpl-1} introduced by \citet{Steiner2009}. 
Later we discuss, how the results would change if \textsc{simpl-2} was used (see Sect. \ref{sec:discussion}).
The input black-body spectrum is convolved with the Green's function describing the scattering \citep[see Equations (1) and (4) in][]{Steiner2009} into a Comptonized power-law-like spectrum.

For a given temperature $T'$ of the spot, we compute a seed black-body spectrum $I'_{\mathrm{bb}}(E')$ at a grid of photon energies logarithmically spaced from 0.1 to 100 keV with a step $\Delta\log E'=0.02$. 
Using then \textsc{simpl}, we obtain also a table of intensities for the Comptonized spectrum $I'_{\mathrm{comp}}(E')$. 
For each emitting point, we compute the intensity as 
\be\label{eq:simplspec}
 I'_{E'}(\sigma ') =  (1-X_{\mathrm{sc}})I'_{\mathrm{bb}}(E')+f (\sigma ')I'_{\mathrm{comp}}(E'),
 \ee
where  the black body is assumed to be isotropic and  the angular dependence of the Comptonized emission is given by the function 
\be\label{eq:beaming}
f(\sigma ') =  I_{0}(1+h\cos\sigma'),
 \ee
 with $h$ being the (free) anisotropy parameter and $I_{0}$ being the normalization, such that 
 $2\int^{\pi/2}_0 \cos \sigma' f(\sigma ')  \sin \sigma' \d  \sigma' = 1$. 

\subsubsection{Disc reflection}\label{sec:ironref}

In order to fit the spectrum of \source, we need also take into account the reflection of the photons from the accretion disc. 
In addition to the Compton reflection continuum that produces spectral hardening above $\sim$10 keV, a fluorescent K$\alpha$ iron line close to 6.4 keV is emitted as observed by {\citet{GDB02}}. 
To describe reflection, we use the \textsc{xspec} model \textsc{xilconv}, which is a convolution model combining reflection model \textsc{xillver} from an ionized disc \citep{GDR13} with the Compton reflection code by \citet{MZ95} to give a more correct representation of Compton recoil effect. 
It is a modified version of the \textsc{rfxconv} model described in \citet{Kolehmainen11}.

We use the \textsc{xilconv} algorithm to re-compute all the energy spectra separately in each of our modelled phase bins. 
The original phase-resolved spectra are given as input. 
We fix most of the additional parameters of the reflection model at the best-fits of the observed phase-average spectrum of \source \ with the model \textsc{phabs}$\times$(\textsc{xilconv}(\textsc{simpl}(\textsc{bbodyrad}))). 
We assume that the redshift parameter is zero, because the accretion disc is not expected to be very close to the compact object. 
We also assume that the exponential cut-off energy is 300~keV. 
From the \textsc{xspec} fits we get that the iron abundance is ten times higher than the solar iron abundance, and that the ionization parameter $\log\xi\approx3.2$. 
The relative reflection normalization $X_{\mathrm{refl}}$ is still kept as a free parameter, as well as the inclination angle $i$ (same angle as in the pulse profile modelling). 

\subsubsection{Response convolution}\label{sec:response}

In this work, the instrumental response of the detector has been taken into account by convolving all modelled physical fluxes with response matrix of the Proportional Counter Array (PCA) at the \textit{Rossi} X-ray Timing Explorer (RXTE) satellite. 
This is done by integrating the product of modelled photon number flux $N(E)$ and response matrix $R(n,E)$ over the energies $E$ to get modelled counts $C_{\mathrm{model}}(n)$ in each observed energy channel $n$. 
The integration is performed from the lowest response energies up to $E_{\max} = 60$~keV (above which the source flux would contribute negligibly to the counts observed in the RXTE/PCA band below $\sim$20~keV) using equation 
\be \label{eq:respconv}
C_{\mathrm{model}}(n)= T_{\rm obs} \int_{0}^{E_{\mathrm{max}}} N(E)R(n,E)\d E , 
\ee
where $T_{\rm obs}$ is the observing time. 
The values of photon flux $N(E)$ are computed in 50 logarithmic energy bins from 1~keV to $E_{\max}$. 
The integration is performed using denser energy grid of the response matrix. 
The corresponding values of $\log N(E)$ are then linearly interpolated to the desired $\log E$.

In addition, the detected instrumental background of \source\ is kept fixed and it is added to the modelled counts.
These predicted counts are then fitted with the observed counts as explained in the following sections. 
However, only the 24 observational energy channels between 3 and 18 keV have been used.
This is based on the best sensitivity of proportional counter units of RXTE and the exclusion of the highest energies where instrumental background dominates for \source\ \citep{HPC08}.

Interstellar photon absorption affecting at the lowest energies has been taken into account by using neutral hydrogen column density for interstellar absorption $N_{\mathrm{H}}$ as one of the free parameters in our model. 
The computed amount of absorption is based on the \textsc{phabs}-function in \textsc{xspec}. 
In practice, this has very little effect because interstellar absorption is severely affecting the spectra only below 1~keV. 
Hence, instead of using the typically more accurate \textsc{tbabs} we can simplify the treatment by adopting the \textsc{phabs} model.
The difference between these models was investigated by fitting the phase-averaged spectrum of \source \ with \textsc{xspec} using both the model \textsc{phabs}$\times$(\textsc{xilconv}(\textsc{simpl}(\textsc{bbodyrad}))) and the model \textsc{tbabs}$\times$(\textsc{xilconv}(\textsc{simpl}(\textsc{bbodyrad}))).
Our fits show that the difference between the fitted parameters with the two models are at most at 1\% level, and the difference in the fitted flux is almost indistinguishable, even at the lowest energy channels of 3~keV.

\subsection{Bayesian modelling}

In order to constrain parameters of NSs we use Bayesian inference combined with the pulse shape and spectral model presented in Sect.~\ref{sec:pulses} and Sect.~\ref{sec:spectra}. 
We fit the model to the observed or synthetic data and use Markov Chain Monte Carlo (MCMC) methods to integrate over the parameter space to find the most probable values for the parameters of the model. 
As a first step, we use synthetic data in order to test our methods and tools. 
The aim of this work is to determine the credible regions for all the parameters in the pulse profile model, given the synthetic and observed waveforms. 
In particular, the credible regions for the mass and radius are obtained by marginalizing over the uncertainties in other parameters. 

We are interested in the probability of the parameters $\textbf{y}$ of the waveform model when the observed waveform is known. 
The probability distribution of the parameters given the data is $p(\textbf{y}|\mathcal{D})$, where $\mathcal{D}$ is the energy- and phase-resolved waveform data. 
According to the Bayes' theorem this (posterior) probability distribution can be obtained from the likelihood of the data, given the parameter values as \citep[see e.g.][]{GS97}
\be \label{eq:bayes}
p(\textbf{y}|\mathcal{D}) \propto p(\mathcal{D}|\textbf{y})p(\textbf{y}) , 
\ee
where  $p(\mathcal{D}|\textbf{y})$ is the likelihood or the probability distribution of the data given the parameters. 
The factor $p(\textbf{y})$ is the prior probability distribution of the parameter values. 
As a first approximation we use uniform priors for most of our parameters (see discussion of the priors in Sect. \ref{sec:discussion}). 
The constant of proportionality is the inverse of the normalization factor, but it is irrelevant when estimating the values of the parameters in a given model. 

\subsubsection{Probabilities}\label{sec:probabilities}

The likelihood of the data for each sample $p(\mathcal{D}|\textbf{y})$ is calculated by fitting the data to the pulse profile model. 
To be more specific, we assume that the probability density of the data counts $C_{\mathrm{data}}$ given the modelled counts $C_{\mathrm{model}}$  is normally distributed around $C_{\mathrm{model}}$ with the square root of the modelled counts as standard deviation similar to Pearson's chi-squared test \citep{cash1979}. 
Adding also intrinsic scatter of the model and the calibration error of the instrument, we have  
\be \label{eq:gaussprob}
p(C_{\mathrm{data}}|C_{\mathrm{model}}) = \frac{\exp \left (\frac{-(C_{\mathrm{data}}-C_{\mathrm{model}})^{2}}{2(C_{\mathrm{model}}+ \sigma_{\mathrm{i}}^{2}+\sigma_{\mathrm{c}}^{2})} \right )}{\sqrt{2\pi(C_{\mathrm{model}}+\sigma_{\mathrm{i}}^{2}+\sigma_{\mathrm{c}}^{2})}},
\ee
where the intrinsic scatter $\sigma_{\mathrm{i}}$ includes the error of the model in describing the data, and $\sigma_{\mathrm{c}} = 0.005 \times C_{\mathrm{model}}$ includes the calibration error of the detector (see Sect. \ref{sec:data}).
In other words, intrinsic scatter is the measure of the systematic errors coming from the choice of the model.
It is a free parameter in the model: if the data are not fully described by the model, it leads to an increase both in $\sigma_{\mathrm{i}}$ and in the credible regions of other parameters.
Normalization is needed because it will not cancel out when calculating likelihood ratios (because of the intrinsic scatter and calibration error). 
In an optimal case, $\sigma_{\mathrm{i}}^{2}$ is very small compared to the count noise term $\sqrt{C_{\mathrm{model}}}$.
The total probability density of the data given the sample is the product of $p(C_{\mathrm{data}}|C_{\mathrm{model}})$ from each phase and energy bin including at least 20 observed photons (or the sum in case of log-probabilities). 

We have also made tests to see that the similar-looking posterior probabilities for pulse profile parameters are also obtained by using Poisson probabilities:
\be \label{eq:poissonprob}
p(C_{\mathrm{data}}|C_{\mathrm{model}}) = \exp (-C_{\mathrm{model}}) \frac{C_{\mathrm{model}} ^{C_{\mathrm{data}}}}{C_{\mathrm{data}}!}.
\ee

The phase shift is treated as a nuisance parameter in our modelling. 
We calculate the probability densities, the products of $p(C_{\mathrm{data}}|C_{\mathrm{model}})$ over phase and energy bins, using different phase shifts. 
A bisection method is used to find the phase shift which has the highest probability. 
As final probability $p(\mathcal{D}|\textbf{y})$ we use the solution with the most probable phase shift. 
Instead of maximizing the likelihood, we could have also marginalized likelihoods over all phase shifts, but the difference between these two methods was found to be marginal.
However, marginalization requires a denser and more evenly distributed phase shift grid than what is possible by using only bisection method (at least in case of very flat light curves).

\subsubsection{Sampling methods}\label{smethods}

To get constraints for our model parameters, we make samples from the posterior distributions using an affine invariant ensemble sampler, which is a MCMC method described in \citet{goodman10}. 
The algorithm has a similar structure to the normal Metropolis scheme and still uses a proposal and either accepts or rejects a step. 
But instead of evolving only one sample value, it evolves an ensemble of sample values, called walkers, together. 
On each iteration the algorithm generates a new sample for every walker using the  current positions of all the other walkers in the ensemble. 
The affine invariant trial move, that we use here, is the so-called stretch move. 
In this method each walker is moved using only one randomly selected complementary walker.

When using the ensemble sampler method to data with significantly less than 1\% errors (and using more than a few free parameters), we encountered problems with walkers either getting stuck into local optima or being unable to move efficiently enough due to the non-linear parameter degeneracies. 
The convergence of the posterior probability distributions was also very slow, even though we used multiple independent ensembles.

To solve these issues, we made few variable transformations (see Sect. \ref{sec:vartr} for more detailed discussion). 
Secondly, we used more precise starting limits for the priors in parameters that could be fitted separately with \textsc{xspec} (spectral parameters) or could be approximated from other studies (size of the spot and distance). 
The recommended technique is to start walkers in a small sphere around the a priori preferred position. 
However, for the radius, mass, observer inclination, and the magnetic inclination, we allowed a larger range of initial positions. 
The exact starting limits ranged from 0.01\% to 20\% of the corresponding widths of the final priors shown in Figs \ref{fig:syntpost}, \ref{fig:syntpostextp}, and \ref{fig:datpost} (with the exceptions mentioned in Sect. \ref{sec:results}). 
The starting points for every walker were drawn randomly from this ball, which was surrounding our initial guess.
For each ensemble, the initial guess was a point obtained by perturbing the correct solution of synthetic data by 4\% of the width of the final priors in each dimension (see Sect. \ref{sec:syntdata} for the correct solution for the synthetic data). 

Finally, we also applied a clustering method described in \citet{Hou12}. 
At certain moments during the burn-in phase of the sampling, the walkers with worst-fitting parameters were drawn to the vicinity of the best-fitting walkers of the same ensemble, if the gap between log-likelihoods of two adjacent walkers (ordered according to decreasing likelihood) became too high compared to differences of other adjacent walkers. 
We also tried to use the simulated annealing method\footnote{
In this method an `artificial temperature' is introduced to create a modified posterior distribution allowing an exploration of parameter space without getting stuck into the local optima. 
The temperature is decreased slowly until the true posterior distribution is obtained. 
However, in our case we did not find any cooling schedule that would be computationally short enough, and would have significantly helped to reduce the multiple solutions found by different ensembles with different starting positions.}
described again in \citet{Hou12}, but found no significant help from that.

In addition, we have checked our results using a multi-modal nested sampling algorithm, implemented in \textsc{MultiNest} \citep{multinest09},\footnote{
https://ccpforge.cse.rl.ac.uk/gf/project/multinest/} 
instead of the ensemble sampler. 
It is a Monte Carlo method targeted at the efficient calculation of Bayesian evidence for a model, but it also produces posterior samples as a by-product. 
The method should have improved efficiency especially for posterior distributions that may contain multiple probability maxima and pronounced degeneracies in high dimensions, that are multi-modal distributions. 
To compute the posterior probability distributions, we used the \textsc{PyMultiNest}\footnote{
https://github.com/JohannesBuchner/PyMultiNest}
package of \textsc{Python}. 

\subsubsection{Variable transformations}\label{sec:vartr}

In all of our models, we have 13 free parameters (in addition to the phase shift). 
The parameters are listed in Table \ref{table:params}. 
To improve the performance of our method, we have used several parameter transformations given as $i+\theta$, $i-\theta$, $M/\req$, $(\req\rho(1+z_{\mathrm{eq}})/D)^{2}$, $(T'/(1+z_{\mathrm{eq}}))^{4}$, $X_{\mathrm{refl}} \cos i$,
and $\log\sigma_{\mathrm{i}}$ as our parameters instead of $i$, $\theta$, $\req$, $\rho$, $T'$, $X_{\mathrm{refl}}$,
and $\sigma_{\mathrm{i}}$, when sampling the parameter space (see discussion in Sect. \ref{sec:anapprox}). 
Here $z_{\mathrm{eq}}$ is the gravitational redshift at $R = \req$.

\begin{table}
  \caption{Parameters of the synthetic data.}
\label{table:params}
\centering
  \begin{tabular}[c]{ l  c } 
    \hline\hline
     Parameter & Value\\ \hline
      Equatorial radius $\req$ & $12.0$ km  \\ 
      Mass $M$ & $1.5 ~\msun$  \\ 
      Inclination $i$ & $60 \degree$ \\ 
      Spot co-latitude $\theta$ & $15 \degree$ \\ 
      Spot angular radius $\rho$ & $15.5 \degree$  \\ 
      Distance $D$ & $3.5$ kpc \\ 
      Temperature $T'$ & $0.85$ keV \\
      Beaming parameter $h$ & $-0.7$  \\
      Scattered photon fraction $X_{\mathrm{sc}}$ & $0.6$  \\
      Photon spectral index $\Gamma$ & $1.8$  \\
      Intrinsic scatter $\sigma_{\mathrm{i}}$ & $0.0$  \\
      Hydrogen column density $ N_{\mathrm{H}}$ & $10^{21} \mathrm{cm}^{-2}$ \\  
      Relative reflection from accretion disc $ X_{\mathrm{refl}}$ & $0.05$ \\
    \hline
  \end{tabular}
  \end{table}

The choice of sampling $i+\theta$ and $i-\theta$ is easy to explain, because the pulse profile is nearly degenerate to switching $i$ and $\theta$ \citep[see e.g.][ and Sect. \ref{sec:anapprox}]{VP04} and  depends only on $x \equiv \sin i \sin \theta$ and $y \equiv \cos i \cos \theta$. 
Therefore, it is easier to get constraints for $x$ and $y$, or, alternatively, for $y+x$ and $y-x$. 
Because $i-\theta = \pm \arccos(x+y)$ and $i+\theta = \arccos(y-x)$, we get better constraints also for $i+\theta$. 
For $i-\theta$ we expect still a bimodal distribution, if there is no prior information of these angles. 
The inclination has been constrained to $i = 36\degr-67\degr$ using optical observations \citep{DHT08}.  
The X-ray analysis of the 2002 outburst  \citep{IP09} led to a similar constraint of $i=50\degr-70\degr$. 
Modelling the broadened iron line  with observations from \textit{Suzaku} and \textit{XMM-Newton}, \citet{CAP09} obtained $i = 51 \degr -63 \degr$ with $90 \%$ confidence. 
In any case, due to the lack of X-ray eclipses, the inclination should be below $82\degree-84\degree$, depending on the assumed NS mass \citep{CM98}. 
Using slightly more conservative values, we first limit the inclination still between $40 \degree$ and $90 \degree$ in our model, in order to be as model-independent as possible.
Later we test, how the tighter constraints on inclination, will affect the results (see Sect.~\ref{sec:add_obs_constr}).

The choices of sampling mass-radius-ratio $M/\req$, normalization parameter $(\req\rho(1+z)/D)^{2}$, the observed temperature in the fourth power $(T'/(1+z))^{4}$, and the angle-corrected relative reflection $X_{\mathrm{refl}} \cos i$ are based on our expectations of which variables most directly affect the observed pulse profiles (see Sect.~\ref{sec:anapprox}). 
These parameter transformations are also aimed to help sampling with for example the coupled $\req$ and $\rho$. 
In the case of $\sigma_{\mathrm{i}}$, we sample $\log\sigma_{\mathrm{i}}$, because the intrinsic scatter is a scale parameter of the model. 
Even though the original non-transformed parameters were not sampled, we show their posterior distributions in Sect.~\ref{sec:results}.

\section{Data}\label{sec:data}

\subsection{SAX J1808.4$-$3658}\label{sec:datareal}
The primary data used in this article is the phase-resolved energy spectrum of \source \ observed during its outburst in 1998 by RXTE. 
The energy-dependent pulse profiles are constructed in the same way as in {\citet{PG03}}. 
We use the data binned in $16$ phase bins and use the $24$ energy channels which extend between $3$ keV and $18$ keV, as justified in Sect.~\ref{sec:response}. 
The data are obtained by combining observations of all the 5 proportional counters of the PCA/RXTE and observations 
between 1998 April 11 and 29 (identical to that  used by \citealt{PG03}).
During this period the pulse shape stayed most of the time almost constant (see Fig. 3 in \citealt{HPC08}).
Therefore, we do not expect large changes in best-fit mass and radius when using a shorter interval.
This was also noted by \citet{ML11}.
The exposure time of each phase bin is approximately 10.5 hours, and the total number of detected counts is close to $4\times10^{7}$. 
This means that the mean statistical uncertainty in our 384 phase-energy-bins is approximately 0.3\%.
The systematic calibration error of PCA is about 0.5\% \citep{JMR06}.
Although, the response files we use were generated in 2003 with a possibly slightly higher calibration error, this not critical, because the underestimated calibration error should be captured by the intrinsic scatter $\sigma_{\mathrm{i}}$ term when fitting the data (see Sect. \ref{sec:probabilities}).

\subsection{Synthetic data}\label{sec:syntdata}

In order to test the method and measure its accuracy, we also apply the fitting framework for the synthetic data, which are generated using our model, and thus we know the result.
The synthetic data in this work are designed to resemble the actual observations of \source\ as closely as possible. 
The amplitude of the modulation in the normalized number fluxes has been set close to the same value in the synthetic data as in the observations. 
We also use the same energy intervals and the same number of phase bins as in the actual data. 
In addition, the total number of counts is the same as in the observed data (see Sect.~\ref{sec:datareal}).

The synthetic data have been generated using our pulse profile model with parameters shown in Table \ref{table:params}. 
In addition, we have fixed the rotational frequency of the star at $401$ Hz, according to the observations of \source \ \citep{WvdK98}.
We have also performed an extra phase shift to our synthetic pulse profile so that the best-fitting phase shift is not immediately found, which is also the case for the observed data.

When creating the data, we have used a higher phase resolution (500 phase bins) in the model than when actually fitting the data (128 phase bins). 
In both cases, the physical fluxes are calculated with 50 energy bins between 1 and 60 keV. 
The synthetic pulse profiles are interpolated and integrated to 16 phase bins (similar to the data of \source) and convolved with the response matrix, in order to get light curves in the 24 PCA energy channels between 3 and 18 keV. 
In order to include the noise, we have Poisson sampled the detected counts at each phase point and energy channel. 
Thus, a relatively larger noise appears at the highest energy channels where the count rates are lowest. 
The 0.5\% calibration error of the PCA instrument is included in to the fit because of the systematic uncertainties in the response matrix (see Sect. \ref{sec:probabilities}).

\section{Results}\label{sec:results}

\subsection{Synthetic data}\label{sec:syntsax}
We begin our analysis with synthetic data created as described in the previous section. In the first model we keep all our parameters free and use the least constraining limits for prior distributions. 
The uniform prior probability is set for $\cos i$ instead of $i$, because we assume the observers to be uniformly distributed on a sphere around the star.
Because $\d \cos i \propto \sin i ~\d i$, we have $\sin i$ prior probability in $i$. 
Otherwise, we use uniform prior distributions for our sampling parameters. 
Of course, since the sampling parameters differ from the model parameters (see Sect.~\ref{sec:vartr}), the prior probabilities in $\req$, $i$, $\theta$, $\rho$, $T'$, and $X_{\mathrm{refl}}$ are multi-dimensional and non-uniform (and therefore not shown in the figures of this article).

\begin{figure*}
\centering
\includegraphics[width=18cm]{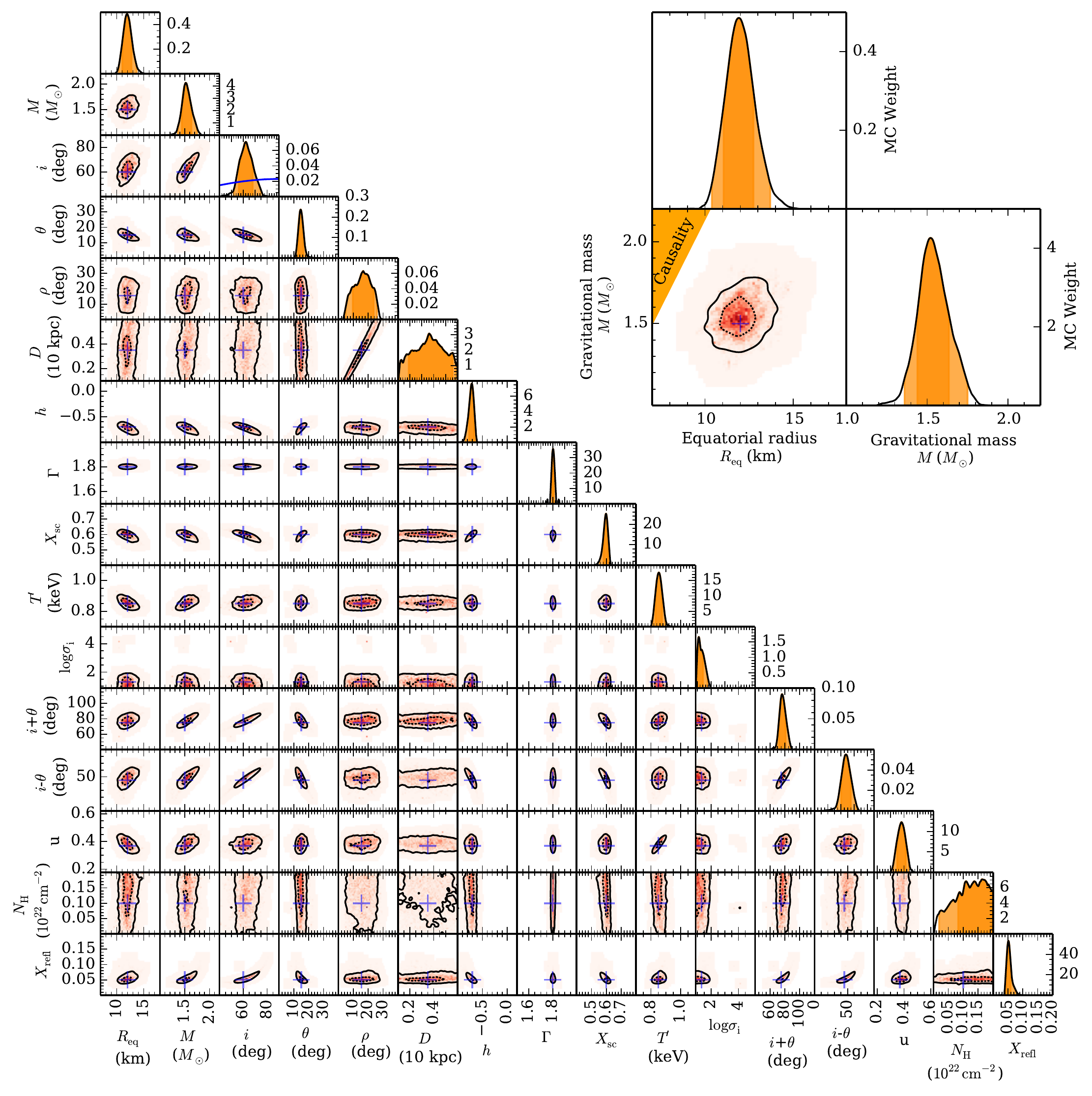}
\caption{Posterior probability distributions for the MCMC runs with the synthetic data. The dark orange colour shows a 68\% and the light orange colour a 95\% highest posterior density credible interval. In the 2D posterior distributions the solid contour shows a 95\% and the dashed contour a 68\% highest posterior density credible region. The blue crosses show the correct solution. The $\sin i$ prior in $i$ is shown with a blue line. The compactness parameter $u$ is defined here as $u \equiv \rg/\req$. The inset in the upper right corner shows the mass-radius posterior distribution in more detail. 
}
\label{fig:syntpost}
\end{figure*}

The resulting posterior distributions are shown in Fig.~\ref{fig:syntpost}. 
The intervals shown in this figure are also the boundaries of the prior distributions we have used, excluding the upper limit $h = 1.0$ for the beaming parameter and $\theta = 90\degr$ for the magnetic obliquity.
Also, $M/\req$, $i-\theta$, and $i+\theta$ are not sampled in the range shown in Fig.~\ref{fig:syntpost}, but in a range allowing all $M$, $\req$, $i$, and $\theta$ be inside their prior, but rejecting the step if any of the parameters would be outside of it.
In addition, we have estimated a more strict upper limit for $M/\req$ from the causality condition $u \equiv 2GM / \req c^{2} \lesssim 0.96 \times 2/3$ \citep{LP04}. 
The factor $0.96$ is added to ensure that the causality condition is valid also at the pole of the oblate NS.
We also limit the spot temperature in the observer frame $T$ to be in the range $0.6-0.7$ keV, which is obtained by spectral fitting with \textsc{xspec}.

\begin{figure}
\resizebox{\hsize}{!}{\includegraphics{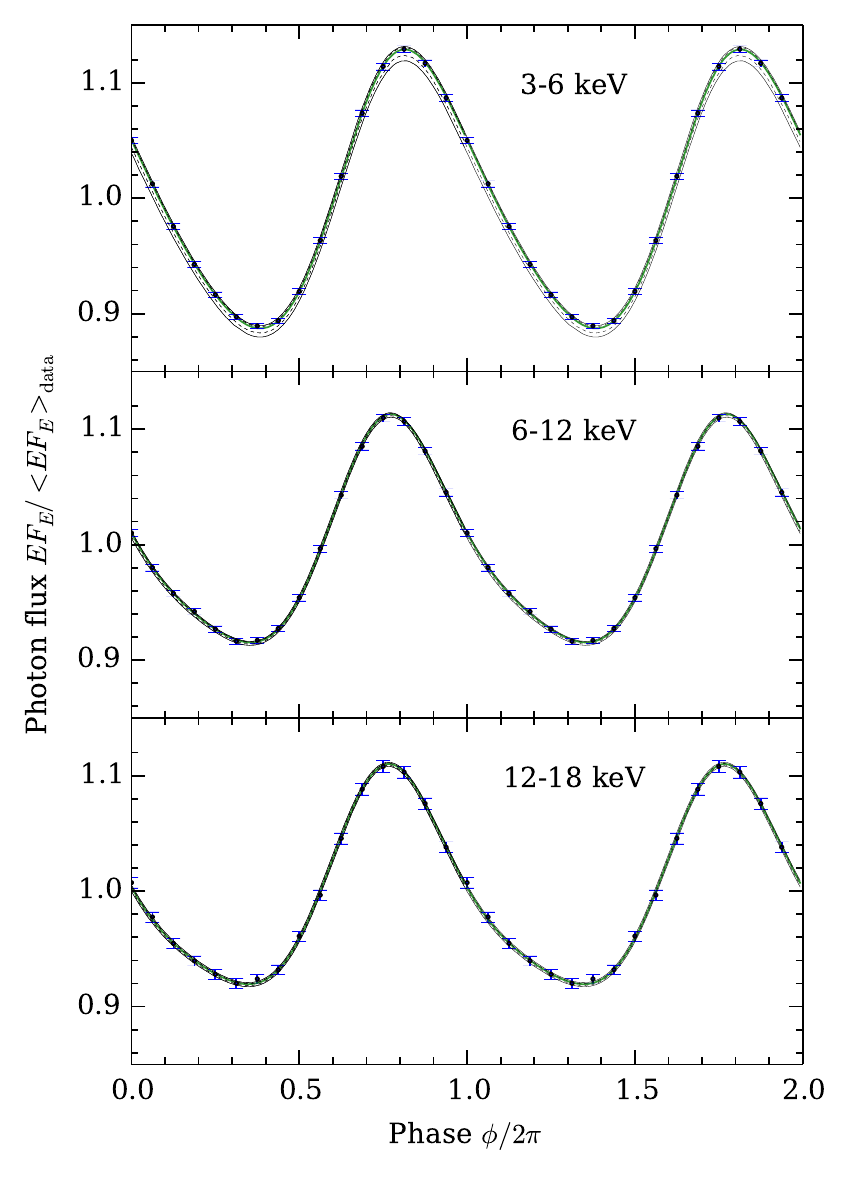}}
\caption{Normalized pulse profiles integrated to 3 energy bins for the synthetic data. The solid black contour shows a 95\% and the dashed black contour a 68\% highest posterior density credible region. The green solid line shows the best-fit solution. The observed data converted to the physical units using the best-fit model are shown with blue circles with the error bars according to the Poisson noise. The 0.5\% calibration error of the detector, used in the fitting procedure, is not included to the error bars.}
\label{fig:syntpulsefit}
\end{figure}

\begin{figure}
\resizebox{\hsize}{!}{\includegraphics{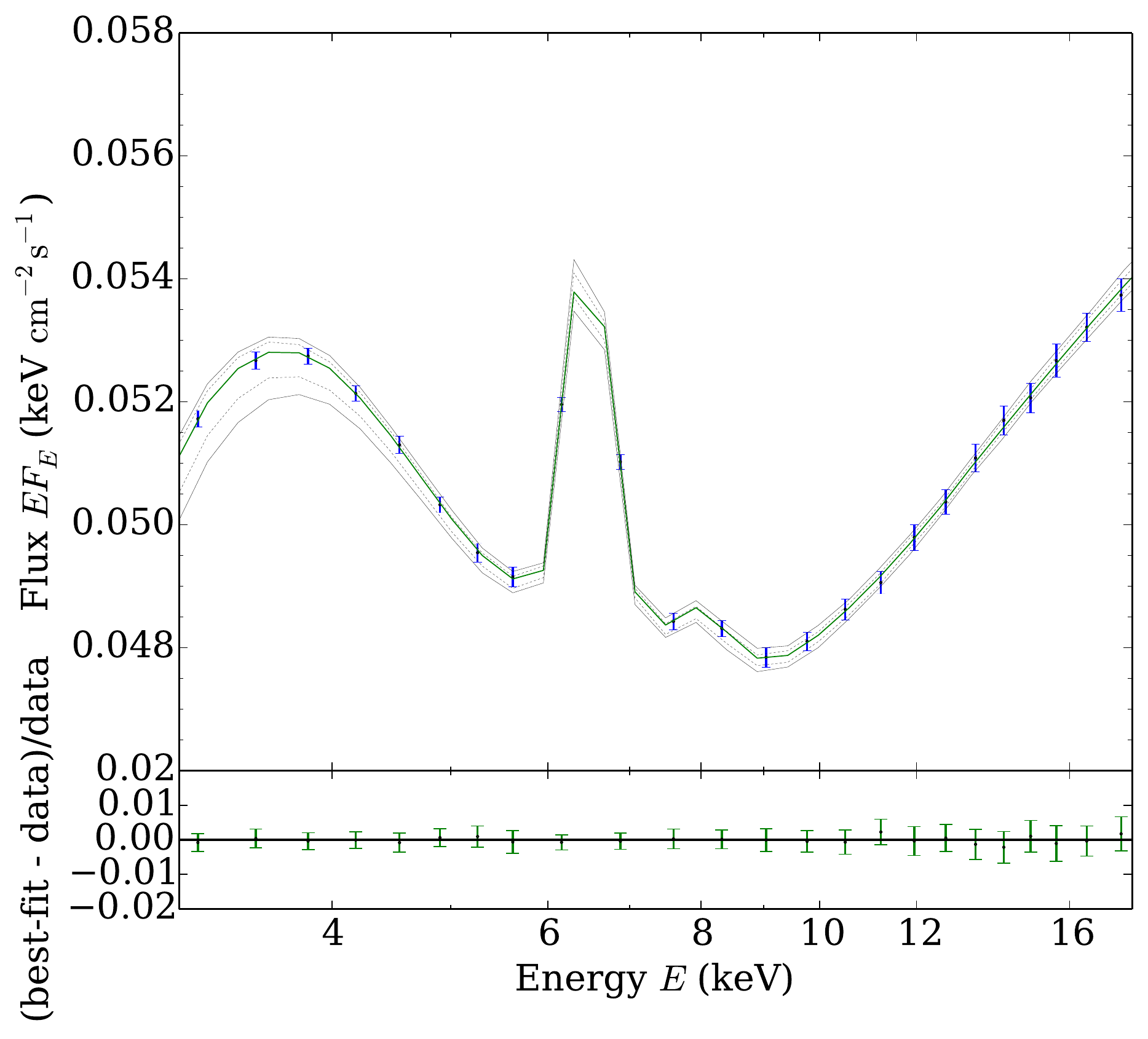}}
\caption{Time-averaged spectrum for the synthetic data. 
The \textit{lower panel} shows the residuals. The contours and the symbols are the same as in Fig.~\ref{fig:syntpulsefit}. The 0.5\% calibration error of the detector, used in the fitting procedure, is not included to the error bars.} 
\label{fig:syntspecfit}
\end{figure}

The contours for pulse profiles (integrated to three energy intervals for easier visualization purposes only) and for the spectrum are shown in Figs.~\ref{fig:syntpulsefit} and \ref{fig:syntspecfit}. 
These figures demonstrate that very good fits are found both in time and energy dimension. 
The best-fit solution presented has $\chi^{2}/\mathrm{d.o.f.} = 119.1/(384-14) \approx 0.32$ (for 14 free parameters including the phase shift).
The reduced chi-squared value is very low, because we have fitted the data using calibration error, even though there is no such error in the observed synthetic counts.  
Without the calibration error we would have $\chi^{2}/\mathrm{d.o.f.} = 375.6/370 \approx 1.02$.
Also, the probability distribution around the best-fit is very tightly concentrated around the correct solution. 

From Fig.~\ref{fig:syntpost} we see that the posterior probability distributions for each parameter are more or less close to the original input values of the synthetic data. 
As expected, the parameter $i-\theta$ is more difficult to constrain than $i+\theta$. 
But from two solutions with opposite signs, we find the correct one, since the solution with switched $i$ and $\theta$ is, fortunately, below our lower limit for inclination, which is $40 \degree$. 
However, the inclination angle $i$ is slightly biased towards higher angles.
On the other hand, $\theta$ is better constrained than $i$.

Most of the parameters of our model seem to be very well constrained. 
The relative reflection $X_{\mathrm{refl}}$, the scattered fraction of photons $X_{\mathrm{sc}}$, the spot temperature $T'$, the photon spectral index $\Gamma$, and the beaming parameter $h$ are the most tightly constrained.  
The angular radius of the spot $\rho$ has a relatively broad probability distribution, however, not as broad as the distance to the star $D$. 
The distribution for the hydrogen column density $N_{\mathrm{H}}$ is also very broad, and it is slightly biased towards higher values.
The uncertainty is, however, expected because $N_{\mathrm{H}}$ will only affect the lowest energy channels.

The most interesting parameters regarding this work are, of course, the mass $M$ and the radius $\req$. 
The results in Fig.~\ref{fig:syntpost} show that even with using only broad 
priors and high level of flexibility in our model, we are able to get meaningful constraints for masses and radii. 
The correct value for the synthetic data is located inside the $68\%$ credible region. 
The probability distribution for the mass seems to slightly favour too high masses. 
We do not see any significant biases in the radius. 
All calculated credible limits and the most probable values are given in Table \ref{table:conflimits}. 
The limits are calculated using the highest posterior probability credible intervals. 
For the radius we find the 68\%  (95\%) limits and the most probable value as $\req = 12.0^{+0.8 ~(1.7)}_{-0.9 ~(1.6)}$ km, and for the mass $M = 1.51^{+0.13 ~(0.24)}_{-0.07 ~(0.15)} ~\msun$. 

The posterior probability distribution for intrinsic scatter $\sigma_{\mathrm{i}}$ shows that the model describes the synthetic data well.
The mean $\log \sigma_{\mathrm{i}} \approx  1$ of the posterior translates to an error of 10--100 counts in each phase bin. 
This can be compared to the level of Poisson noise in the data, which is approximately $300$ counts on average in each phase bin.
Therefore, $\log \sigma_{\mathrm{i}} \approx  1$ is effectively the same as $\sigma_{\mathrm{i}} \approx  0$, as expected for synthetic data.



\subsection{Synthetic data with additional polarimetric constraints}

\begin{figure*}
\centering
\includegraphics[width=18cm]{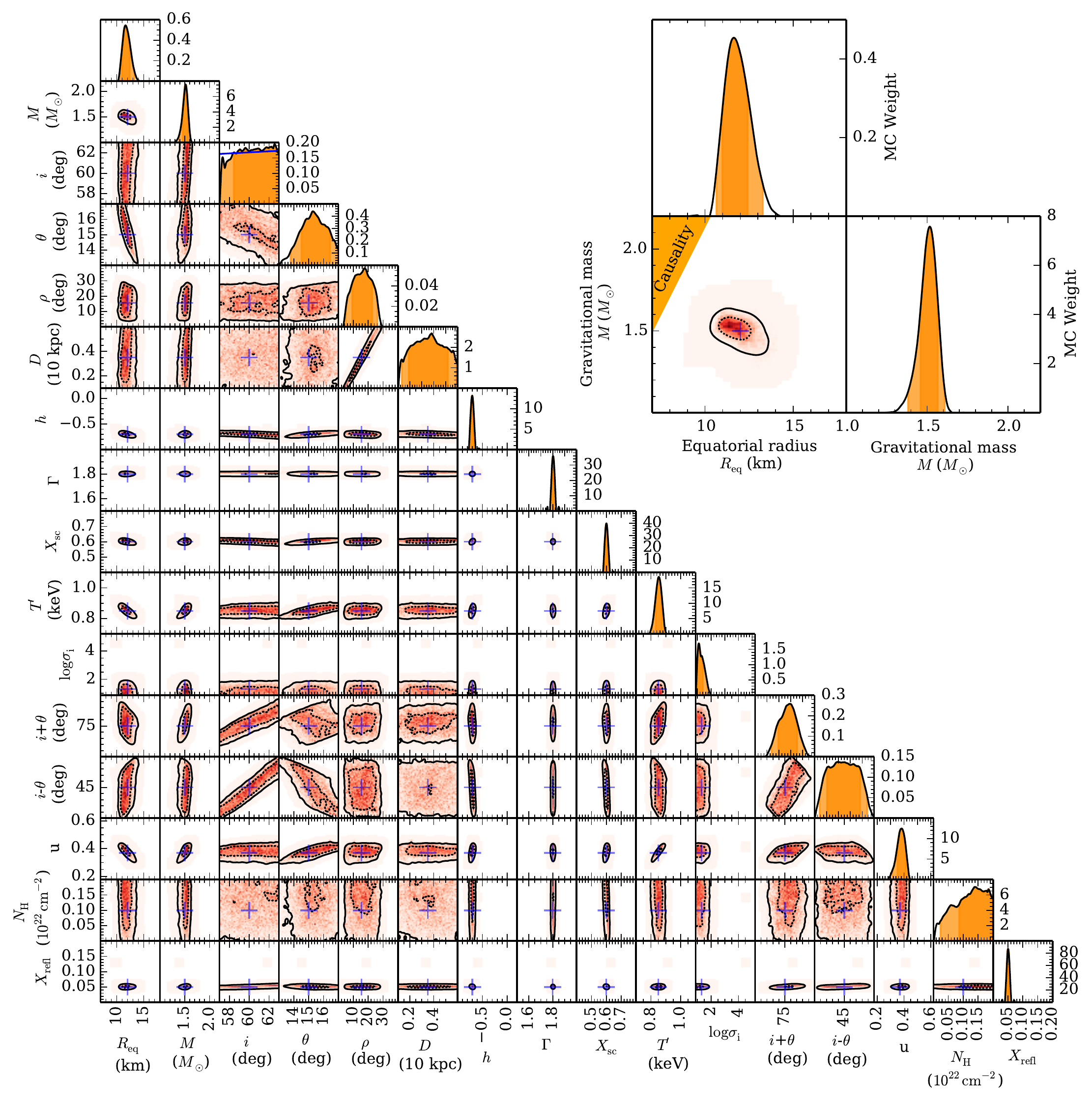}
\caption{Posterior probability distributions for the MCMC runs with the synthetic data, where angles $i$ and $\theta$ are better constrained using the data possibly coming from IXPE. The colours, contours, and other symbols are the same as in Fig.~\ref{fig:syntpost}.
}
\label{fig:syntpostextp}
\end{figure*}

In this section we show what is the effect of having prior information of observer inclination $i$ and spot co-latitude $\theta$. 
This information could be obtained using the polarization measurements from, for example, the upcoming IXPE mission. 
The swing of the polarization angle with phase can be used to determine both $i$ and $\theta$ \citep{VP04}. 
Our synthetic data are exactly the same as in Sect.~\ref{sec:syntsax}.
However, we assume that the polarization data allow constraining inclination between $57 \degr$ and $63 \degr$ and the spot co-latitude between $13 \degr$ and $17 \degr$ (i.e. around the assumed parameters, see Table~\ref{table:params}), restricting therefore the prior distributions. 
The boundaries of the prior distributions for other parameters  are the same as in Sect.~\ref{sec:syntsax}.
The results are shown in Fig.~\ref{fig:syntpostextp}.
We emphasize here that the values used here might not be exactly correct but they do provide a rough estimate of how the possible polarization measurements might improve our inference.

From the resulting posterior distribution, we see that mass and radius, among all the other parameters, are only slightly better constrained when compared to Fig.~\ref{fig:syntpost}. 
The biases in $i$ and $N_{\mathrm{H}}$ point to the same direction as previously.
The resulting credibility limits can be again found from Table \ref{table:conflimits}. 
In this case, we get a radius $\req = 11.9^{+0.5 ~(1.4)}_{-0.9 ~(1.3)}$~km, and for the mass $M = 1.53^{+0.04 ~(0.08)}_{-0.07 ~(0.15)} ~\msun$.
Compared to the previous model, additional constraints for the geometrical angles improved the accuracy clearly more in mass than in the radius.
The accuracy in determination of mass, with 68\% credibility level, increased from 7\% to 4\%, when the corresponding accuracy for radius increased from 8\% to 6\%. 
For the 95\% credibility level, the accuracy in mass improved from 13\% to 8\% and in the radius from 14\% to 12\%.

\subsection{SAX J1808.4$-$3658 with RXTE}\label{sec:datarealfit}

\begin{figure*}
\centering
\includegraphics[width=18cm]{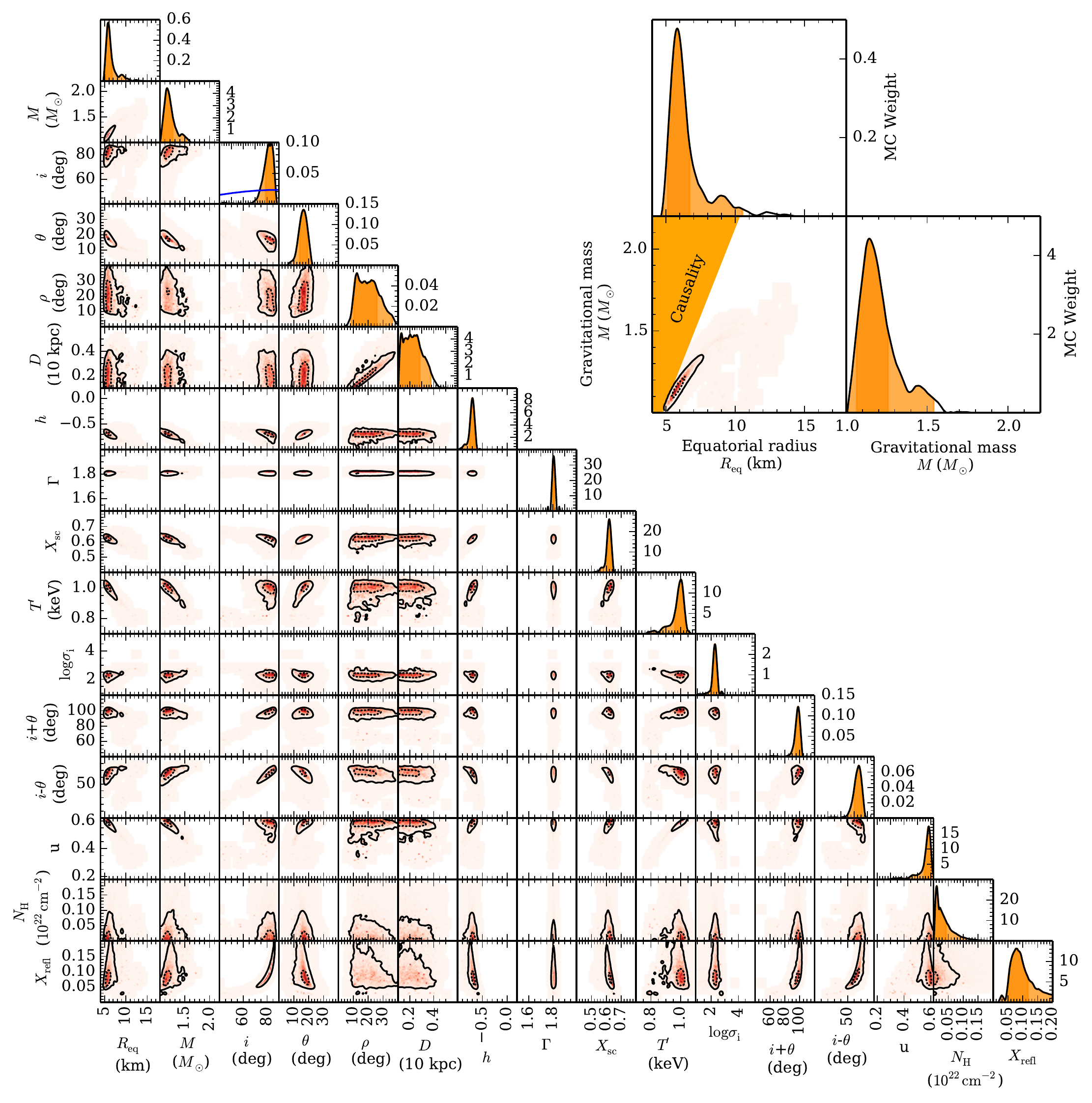}
\caption{Posterior probability distributions for the MCMC runs with \source\ data. The colours, contours, and other symbols are the same as in Fig.~\ref{fig:syntpost}.
The posterior shows that letting the mass be a free parameter results in unrealistic radius estimates due to possible biases in other parameters when the model does not describe data perfectly. 
More realistic parameter constraints with a fixed mass grid are shown in Fig. \ref{fig:datpost_fm_all}.
}
\label{fig:datpost}
\end{figure*}

We now analyse the actual data of \source\ from the 1998 outburst obtained with the RXTE. 
We use the same model as in Sect. \ref{sec:syntsax} to fit the data. 
The resulting posterior probability distributions 
are shown in Fig.~\ref{fig:datpost}. 
The limits of the priors are the same as in Sect.~\ref{sec:syntsax}, except the lower limit for the radius is decreased to 4 km.

\begin{figure}
\resizebox{\hsize}{!}{\includegraphics{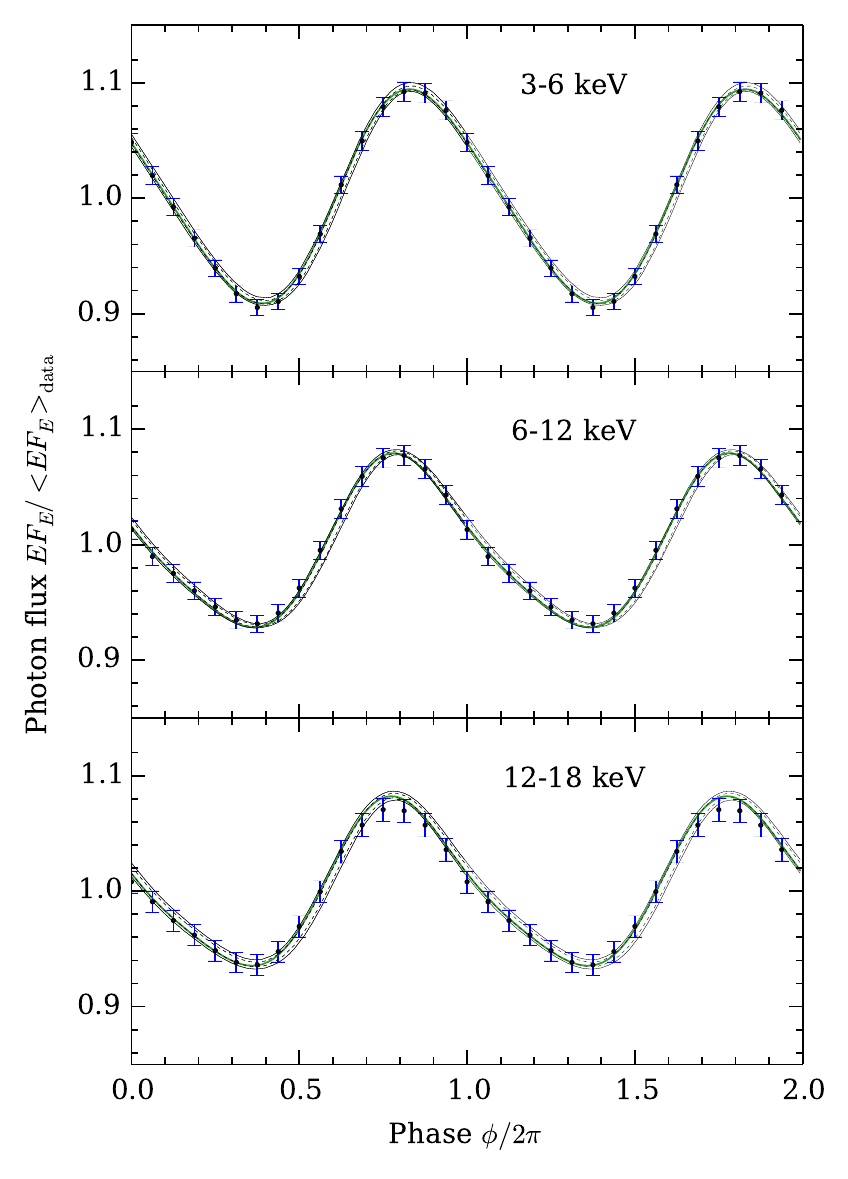}}
\caption{Normalized pulse profiles integrated to 3 energy bins for \source\ data. The contours and the symbols are the same as in Fig.~\ref{fig:syntpulsefit}. 
The 0.5\% calibration error of the detector is included to the error bars.}
\label{fig:datpulsefit}
\end{figure}

\begin{figure}
\resizebox{\hsize}{!}{\includegraphics{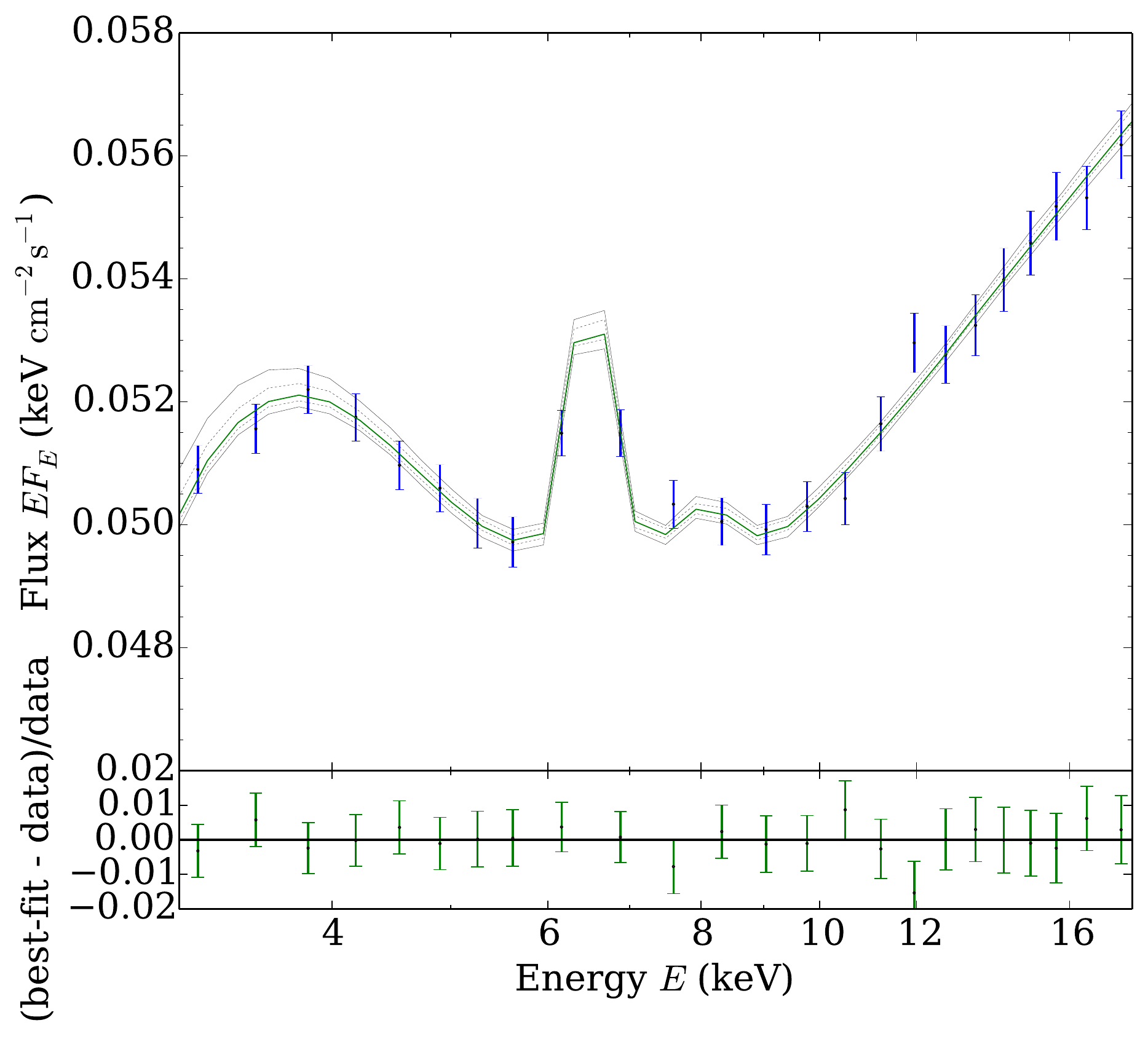}}
\caption{Time-averaged spectrum for \source. 
The contours and the symbols are the same as in Fig.~\ref{fig:syntpulsefit}. The 0.5\% calibration error of the detector is included to the error bars.}
\label{fig:datspecfit}
\end{figure}

The contours for the pulse profiles and for the spectrum are again shown in Figs.~\ref{fig:datpulsefit} and \ref{fig:datspecfit}. 
From these figures we see that the reasonably good fits are found both in time and in energy dimensions. 
The best-fit solution presented has $\chi^{2}/\mathrm{d.o.f.} = 420.6/370 \approx 1.14$ where we have taken into account the 0.5\% calibration error of the PCA. 
Also, the probability distribution around the best-fit is quite tightly concentrated around the observed data. 
However, at the higher energy channels the fits are worse and they are not fully able to produce the observed smaller pulse amplitude (see the energy band 12--18~keV in Fig. \ref{fig:datpulsefit}).

From the posterior distributions we can see that many of the parameters of our model are quite well constrained. 
For example, the photon spectral index $\Gamma$ and the scattered fraction of photons $X_{\mathrm{sc}}$ are reasonably well-constrained close to the values they should have according to the independently performed spectral fits with \textsc{xspec} ($\Gamma = 1.82^{+0.04}_{-0.03}$  and $X_{\mathrm{sc}} = 0.60^{+0.02}_{-0.05}$ with $90 \%$ confidence ranges).  
The beaming parameter $h = -0.69^{+0.04 ~(0.07)}_{-0.06 ~(0.20)}$ is constrained close to the results of \citet{PG03}, and the reflection amplitude $X_{\mathrm{refl}} = 0.09^{+0.03 ~(0.11)}_{-0.04 ~(0.06)}$ is consistent with the results of \citet{GDB02}. 
The posterior of the hydrogen column density $N_{\mathrm{H}}$ peaks at a value factor of twenty lower than expected. 
However, this is not critical because it just implies that the data do not contain enough information to put any stringent constraints on it.
Later we confirm this by fixing the value of $N_{\mathrm{H}}$  to  $0.14 \times 10^{22} ~\mathrm{cm^{-2}}$ given by \citet{PKC13} (see Sect. \ref{sec:add_obs_constr}).

The angular size of the spot $\rho$ has a relatively broad probability distribution, like in the case of the synthetic data, but it reaches slightly higher values (up to about $30 \degree$).
For the distance we get $D = 2.5^{+0.3 ~(1.3)}_{-1.4 ~(1.5)}$ kpc, where the most probable distances are slightly smaller than the most probable distance of 3.5~kpc given by \citet{galloway2006}. 
The source inclination is constrained to lie above 71\degr.
 
The credibility limits for parameters, when fitting the observed data, are also shown in Table \ref{table:conflimits}. 
In case of radius $\req$, our original choices to the limits of the priors would have strongly constrained the results. 
In this case, we get a radius $\req = 5.8^{+0.9 ~(4.7)}_{-0.6 ~(1.0)}$ km, and for the mass $M = 1.13^{+0.13 ~(0.41)}_{-0.06 ~(0.12)} ~\msun$. 
These results favour smaller NSs than generally accepted. 
However, this could occur, for example, due to any biased estimates in the other model parameters (see Sect.~\ref{sec:discussion} for more detailed discussion).

The posterior probability distribution for intrinsic scatter with the mean $\log \sigma_{\mathrm{i}} \approx 2.3$ shows that the model does not describe the data as well as the synthetic data. 
This corresponds an absolute error of about 200 counts or a relative error of 0.2\% in each phase-energy bin. 
However, this error is still smaller than the Poisson noise level in the data (see Sect. \ref{sec:data}).

\begin{table*}
  \caption{Most probable values and $68\%$ and $95\%$ credible limits for all 3 different simulations.}
\label{table:conflimits}
\centering
  \begin{tabular}[c]{l c c c c c}
    \hline\hline
      Quantity & $95\%$ lower limit & $68\%$ lower limit & Most probable value& $68\%$ upper limit & $95\%$ upper limit \\ \hline
     \multicolumn{6}{c}{Synthetic data} \\ 

      $\req$ (km) & $10.4$ & $11.1$ & $12.0$ & $12.8$ & $13.7$  \\ 
      $M$ ($\msun$) & $1.36$ & $1.44$ & $1.51$ & $1.64$ & $1.75$  \\ 
      $i$ ($\deg$) & $52$ & $56$ & $61$ & $68$ & $75$ \\ 
      $\theta$ ($\deg$) & $12$ & $13$ & $15$ & $16$ & $18$ \\ 
      $\rho$ ($\deg$) & $4.8$ & $9.7$ & $16$ & $24$ & $26$ \\ 
      $D$ (kpc) & $1.1$ & $1.9$ & $3.6$ & $5.6$ & $5.9$ \\ 
                                                                
      $h$  & $-0.84$ & $-0.77$ & $-0.70$ & $-0.66$ & $-0.63$ \\ 
      $T'$ (keV) & $0.81$ & $0.83$ & $0.85$ & $0.87$ & $0.90$ \\ 
      $X_{\mathrm{sc}}$ & $0.56$ & $0.58$ & $0.60$ & $0.61$ & $0.62$ \\
      $\Gamma$ & $1.794$ & $1.798$ & $1.803$ & $1.807$ & $1.812$ \\   
      $\log \sigma_{\mathrm{i}}$ & $0.90$ & $0.92$ & $1.05$ & $1.42$ & $1.75$ \\ 
      $N_{\mathrm{H}}$ ($10^{22} \mathrm{cm}^{-2}$) & $0.019$ & $0.083$ & $0.14$ & $0.20$ & $0.20$ \\
      $ X_{\mathrm{refl}}$ & $0.040$ & $0.044$ & $0.051$ & $0.059$ & $0.076$ \\   \hline
                 
      \multicolumn{6}{c}{Synthetic data with more precise priors for $i$ and $\theta$} \\ 

      $\req$ (km) & $10.6$ & $11.0$ & $11.9$ & $12.4$ & $13.3$  \\ 
      $M$ ($\msun$) & $1.38$ & $1.46$ & $1.53$ & $1.57$ & $1.61$  \\ 
      $i$ ($\deg$) & $57.1$ & $58.8$ & $62.6$ & $62.9$ & $62.9$ \\ 
      $\theta$ ($\deg$) & $13.8$ & $14.6$ & $15.2$ & $16.5$ & $16.9$ \\ 
      $\rho$ ($\deg$) & $4.7$ & $9.2$ & $18$ & $23$ & $26$ \\ 
      $D$ (kpc) & $1.1$ & $1.9$ & $3.7$ & $5.3$ & $5.9$ \\

      $h$  & $-0.75$ & $-0.72$ & $-0.70$ & $-0.68$ & $-0.67$ \\ 
      $T'$ (keV) & $0.81$ & $0.83$ & $0.86$ & $0.87$ & $0.89$ \\ 
      $X_{\mathrm{sc}}$ & $0.59$ & $0.59$ & $0.60$ & $0.61$ & $0.61$ \\
      $\Gamma$ & $1.794$ & $1.798$ & $1.803$ & $1.808$ & $1.812$ \\   
      $\log \sigma_{\mathrm{i}}$ & $0.90$ & $0.93$ & $1.1$ & $1.4$ & $1.7$ \\ 
      $N_{\mathrm{H}}$ ($10^{22} \mathrm{cm}^{-2}$) & $0.025$ & $0.087$ & $0.14$ & $0.20$ & $0.20$ \\
      $ X_{\mathrm{refl}}$ & $0.046$ & $0.048$ & $0.050$ & $0.052$ & $0.055$ \\   \hline
      
      \multicolumn{6}{c}{Observed data of \source} \\ 

      $\req$ (km) & $4.80$ & $5.13$ & $5.77$ & $6.71$ & $10.5$  \\ 
      $M$ ($\msun$) & $1.01$ & $1.07$ & $1.13$ & $1.26$ & $1.54$  \\ 
      $i$ ($\deg$) & $71$ & $79$ & $82$ & $86$ & $87$ \\ 
      $\theta$ ($\deg$) & $9.5$ & $14$ & $17$ & $19$ & $21$ \\ 
      $\rho$ ($\deg$) & $8.4$ & $10$ & $12$ & $26$ & $36$ \\ 
      $D$ (kpc) & $1.0$ & $1.1$ & $2.5$ & $2.8$ & $3.8$ \\  

      $h$  & $-0.89$ & $-0.75$ & $-0.69$ & $-0.65$ & $-0.62$ \\ 
      $T'$ (keV) & $0.79$ & $0.96$ & $1.0$ & $1.0$ & $1.1$ \\ 
      $X_{\mathrm{sc}}$ & $0.56$ & $0.60$ & $0.62$ & $0.63$ & $0.65$ \\
      $\Gamma$ & $1.798$ & $1.803$ & $1.807$ & $1.812$ & $1.816$ \\   
      $\log(\sigma_{\mathrm{i}})$ & $1.65$ & $2.12$ & $2.30$ & $2.44$ & $2.79$ \\ 
      $N_{\mathrm{H}}$ ($10^{22} \mathrm{cm}^{-2}$) & $0.0021$ & $0.0027$ & $0.0049$ & $0.042$ & $0.11$ \\
      $ X_{\mathrm{refl}}$ & $0.027$ & $0.052$ & $0.087$ & $0.12$ & $0.20$ \\   \hline

  \end{tabular}
\tablefoot{The quantities shown in the Table are equatorial radius $\req$, mass $M$, inclination $i$, spot co-latitude $\theta$, spot angular size $\rho$, distance $D$, beaming parameter $h$, temperature of the spot $T'$, scattered photon fraction $X_{\mathrm{sc}}$, photon spectral index $\Gamma$, intrinsic scatter $\log \sigma_{\mathrm{i}}$, hydrogen column density $N_{\mathrm{H}}$, and the reflection amplitude $X_{\mathrm{refl}}$.}  
\end{table*}

\subsection{SAX J1808.4$-$3658 with fixed $M$}

To get more reasonable estimates for the NS radius, we have also computed the posterior distributions for all the parameters with a fixed grid of NS masses from 1.4 to 2.2$~\msun$ with the step of 0.1$~\msun$.
These are then combined to obtain 2D-posterior distributions for every parameter against mass.
The assumption is that each of the mass bins has an equal probability. 
This way it is possible to see what would be the constraints, especially for the radius, if the requirement for the free mass is relaxed and we assume it is to be known instead.

\begin{figure*}
\centering
\includegraphics[width=17cm]{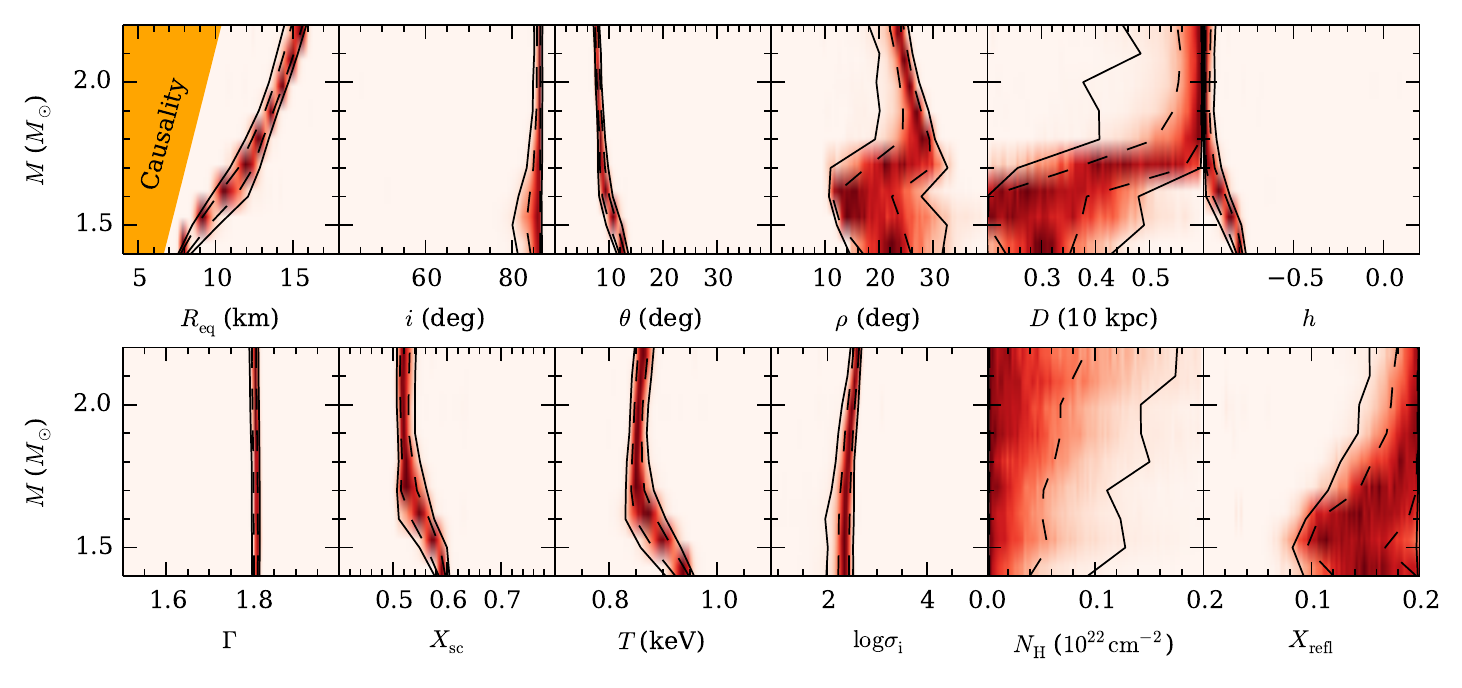}
\caption{Posterior probability distributions for the MCMC runs with \source\ data, where a fixed mass grid has been used. One-dimensional posterior histograms are presented at the top of each other to make approximate two-dimensional histograms with respect to mass. The colours and contours are the same as in Fig. \ref{fig:syntpost}.}
\label{fig:datpost_fm_all}
\end{figure*}

The results for this analysis are shown in Fig.~\ref{fig:datpost_fm_all}.
From the posterior distributions of the intrinsic scatter, we see that all the mass bins give more or less similar fits.  
Most of the parameters depend on the mass only slightly. 
This then indeed implies that the data do not contain enough information to convincingly constrain the radius and the mass simultaneously.
However, some correlation with the mass exists especially in the radius, inclination, distance, temperature, and relative reflection amplitude. 
This time we can get reasonable estimates for the radius; for example, for a NS mass of $1.7~\msun$ we get $\req = 11.9^{+0.5 ~(1.0)}_{-0.4 ~(1.0)}$ km for the equatorial radius.
Furthermore, at this mass we have the most probable distance to the source closest to distance of 3.5~kpc given by \citet{galloway2006}.

\section{Discussion}\label{sec:discussion} 

The results presented in Sect. \ref{sec:results} demonstrate that obtaining reliable radius constraints is difficult without having a priori knowledge of the NS mass.
Our method works well with the synthetic data, but in case of \source\ our model describes the data less accurately, which might lead to larger biases in parameter estimates.
Although, using a fixed grid of NS masses, we obtained more realistic results for the radius, in addition to more likely values for the distance of the star.

When using synthetic data and having additional information on the observer inclination and spot co-latitude, we get only slightly better constraints. 
However, it is not yet exactly known how much the upcoming polarization measurements could limit the possible values of these angles. 
In our case, for the synthetic data, constraining $i$ in the interval $[57 \degree, 63 \degree]$ and $\theta$ in $[13 \degree, 17 \degree]$, instead of $[40 \degree, 90 \degree]$ and $[0 \degree, 90 \degree]$, respectively, resulted in a change of the 68\% credibility interval for the radius from 11.1--12.8 to 11.0--12.4~km and the interval for the mass from 1.44--1.64 to 1.46--1.57~$\msun$. Similarly, the 95\% credibility intervals have changed from 10.4--13.7 to 10.6--13.3~km and from 1.36--1.75 to 1.38--1.61~$\msun$. 
In addition, the most probable posterior value for the mass was found to be closer to the correct solution.

The more significant accuracy improvement on the mass, rather than on the radius, results from the stronger correlation between mass and inclination, than between radius and inclination.
This is also seen in the posterior probability distributions in Figs.~\ref{fig:syntpost} and \ref{fig:syntpostextp}.
From the latter figure, we see that the equatorial radius $\req$ is more correlated with the spot co-latitude $\theta$, as expected for the oblate star.
However, our new limits for $i$ and $\theta$ bind the original posterior more for $i$ than for $\theta$.

The constraints in mass and radius appeared to be very good in both cases where we fitted our synthetic data. 
However, as expected, the obtained constraints were weaker in the case of the observed data. 
This shows that our model does not describe the observations perfectly. 
On the other hand, the most probable values obtained for the radius (from 5 to 11~km), are somewhat lower than usually predicted for NSs. 
However, these results are quite similar than what has previously been obtained by {\citet{PG03}} and {\citet{LMC08}} for the same source \source. 
Of course, the (equatorial) radius seems to depend strongly on the combination of many other parameters, including at least the mass, spot co-latitude, magnetic inclination, spot angular size, distance, beaming, and the spot temperature. 
An error or bias in any of these parameters would give incorrect results in $\req$.
For example, a higher mass than the best-fitted 1.13~$\msun$, would give higher radii, if the compactness $M/ \req$ remained the same.
The results with a fixed mass grid, shown in Fig. \ref{fig:datpost_fm_all}, also support this and give reasonable radius estimates if the mass is known a priori.

The beaming parameter $h$ turned out be better constrained than the mass and the radius, both with synthetic and real data. 
However, the linear $I'_{\mathrm{non-th}} \propto I_{0}  (1+h \cos \sigma')$ model for the beaming might not be the optimal one, because the largest deviations between the model and the data in the pulse profile are found from the reduced pulse amplitude at highest energies (see Fig.~\ref{fig:datpulsefit}).  
The beaming parameter $h$ is expected to be larger in absolute value at higher energies where the non-thermal radiation dominates, and this should flatten the pulse profiles.
The current model may not do that efficiently enough. 

\subsection{Analytical approximations}
\label{sec:anapprox}

In order to understand the relations between the parameters of our model and the observed quantities, we next present approximate analytical expressions for the pulse amplitudes.
These amplitudes are expected to be the best constrained quantities from the observations.
Using these expressions, we may also better interpret our results.

First we  would like to get an estimate for the normalization of the light curve.
This can be done by using the bolometric version of Eq.~\eqref{eq:fluxspot2}, where instead of $(1-u)^{1/2}\delta^{4}I'_{{E}'}(\sigma')$ we have $(1-u)\delta^{5}I'(\sigma')$. 
Assuming a slowly rotating ($\delta \approx 1$) spherical star ($R = \req$), with an always visible small spot ($\d S'\approx \pi (\rho \req)^{2}$) radiating black-body flux ($I'(\sigma') \propto T'^{4}$), the observed normalization, or the phase-independent part of the flux, scales as
\be \label{eq:amp0}
\mathcal{N}_{0} \propto \left( \frac{\rho \req (1+z_{\mathrm{eq}})}{D} \right) ^{2} \left( \frac{T'}{1+z_{\mathrm{eq}}}\right) ^{4},
\ee
where the first factor is proportional to the solid angle the spot occupies on the sky and the second factor is related to the bolometric black-body flux, both in the frame of an observer.
Because linear correlations are the optimal relations between different variables when using ensemble sampler, we chose to sample the solid angle the spot is seen instead of the spot angular size $\rho$, and the bolometric black-body flux instead of temperature $T'$ (see Sect.\ref{sec:vartr}).

We can also get a useful approximation for the bolometric pulse amplitude. 
This time we use the same assumptions as when deriving $\mathcal{N}_{0}$, but we, in addition, allow the radiation to be anisotropic and allow a larger spot size.
This is done by replacing $I_{0}$ with $I_{0}(1+h\cos \alpha')$ in the derivation of oscillation amplitude in the paper by \citet{PG03}.
In case of using the following notations: 
\begin{align}
U &\equiv (1-\rg/\req)\sin i \sin \theta, \\
Q &\equiv \rg/\req + (1 - \rg/\req)\cos i \cos \theta, \\
P &\equiv \frac{\rg}{\req}\tan^{2}\frac{\rho}{2},
\end{align}
we get for the flux \citep[see also][]{PB06}: 
\begin{eqnarray} \label{eq:aprox_puls_prof}
F(\phi) & \propto&  Q + hQ^{2}+\frac{hU^{2}}{2} + hQP + P \nonumber \\ 
&+&  U(1+2hQ + hP)\cos \phi + h \frac{U^{2}}{2} \cos 2 \phi.
\end{eqnarray}
From this we see that the pulse amplitude of fundamental increases with a positive $h$, and decreases with a negative one.

Using Eq. \eqref{eq:aprox_puls_prof}, we get the relative pulse amplitude
\be \label{eq:amp1}
\mathcal{A}_{1} = \frac{U(1+2hQ+hP)}{Q+hQ^{2}+hU^{2}/2 +hQP+P}, 
\ee
where we ignored the harmonic, which has a smaller amplitude.
Without beaming ($h = 0$) and with a small spot ($P=0$), this expression reduces to the usual $\mathcal{A}_{1} = U/Q$ \citep{b02,PG03}. 
This shows the problem with distinguishing $i$ and $\theta$, when fitting the data without independent knowledge of any of the two parameters. 
As explained in the Sect.~\ref{sec:vartr}, we may reduce the number of (almost) indistinguishable parameters by sampling $i+\theta$ and $i-\theta$, instead of $i$ and $\theta$.

In addition, Eq.~\eqref{eq:amp1} and Eq.~\eqref{eq:amp0} show that $\rho$ is involved both in the normalization and the pulse amplitude.
Therefore, even our transformed spot size parameter (see Sect.~\ref{sec:vartr}) that have a linear relation to $\mathcal{N}_{0}$, is not linear with respect to $\mathcal{A}_{1}$ or the observing angles.
Also, the beaming parameter $h$ is coupled with $\rho$.
The pulse amplitude $\mathcal{A}_{1}$ can be also used to predict the accuracy in measurement of NS compactness when the uncertainties in the other parameters of the expression are known and their values estimated \citep[see, e.g.][]{OPA16}.

The ratio between the amplitudes of second and first harmonic is again obtained from Eq. \eqref{eq:aprox_puls_prof} 
\be \label{eq:amp2_aniso}
\mathcal{A}_{2}= \frac{hU/2}{1+2hQ+hP}.
\ee
With a small spot, this reduces to $\mathcal{A}_{2} \propto h \sin i \sin \theta$ \citep{PB06}.
The beaming parameter $h$ is thus strongly connected to the observing angles also through this expression. 
However, the effect of fast rotation (Doppler effect) causes slightly different amplitudes, and introduces a phase shift between the fundamental and the harmonic, skewing the pulse profile and making the relations between parameters even more complex. 

\subsection{Uncertainties and systematic errors}

The possible systematic errors can emerge either from the sampling method or from the physical modelling of the energy-resolved pulse profiles.  
As described in Sect. \ref{smethods}, sampling the parameter space introduces difficulties when the model parameters are highly degenerate and have non-linear relations with each other. 
This issue has been addressed by making variable transformations in order to sample parameters that are easier to constrain observationally and that would have simpler relations with each other.

The method might still suffer from a very slow convergence. 
The posterior distributions were slowly changing even after tens of thousands of samples from individual ensembles, or more than a thousand accepted steps on average for an individual walker. 
The problems with convergence of affine invariant ensemble sampler in high dimensions have been noted by {\citet{huijser2015}}, and the problem is expected to be worse for strongly correlated and complex models.
Our solution was to use very long runs with ensemble sampler, and confirm the results with the independent nested sampling method \textsc{multinest}.
In \textsc{multinest} we did not use any particular initial position for sampling, which also reduced the risk of the results being dependent on the initial position of an ensemble.

In the case of fitting the \source\ data, a systematic error can arise from the assumptions made in physical modelling. 
As already discussed, the angular distribution of the radiation might need a more realistic model. 
Overall, to explain all the spectral features, an accurate and physical atmosphere model of the accreting NS is required.
In future work, we plan to construct the model for the accretion shock and compute self-consistently the angular distribution of the escaping radiation. 
This first-principle model would significantly reduce the number of free parameters and would allow to get better mass-radius constraints. 

The use of model \textsc{simpl-1}, which includes only the up-scattered photons, instead of \textsc{simpl-2} including both up- and down-scattered photons, can be one source of error. 
For non-relativistic thermal Comptonization some part of the photons is also scattered to lower energies. 
The use of \textsc{simpl-1} could thus result in a slightly too small scattering fraction when compensating the down-scattering effects.
It could also explain the unexpectedly low values of hydrogen column density $N_{\mathrm{H}}$, since both the lower $N_{\mathrm{H}}$ and the down-scattering result in a higher number of photons in the lowest energy bands.
However, the mass and radius estimates are not very sensitive to a small change in these parameters.
This was also confirmed by the fit of the \source\ data, where we switched to \textsc{simpl-2}, while keeping all parameters still free.
The only major difference we found was in $X_{\mathrm{sc}}$, which was increased from $0.6$ to $0.7$.

As mentioned in Sect. {\ref{sec:methods}}, we ignore the radiative transfer processes of photons through an accretion column. 
For a typical luminosity of $5\times10^{36}$~erg~s$^{-1}$ \citep{GR98}, the assumed accretion efficiency of 0.15 and the circular radius of the spot of 2~km, the Thomson optical depth through the column is about a unity \citep[see sect.~5.3 of][]{PG03}.
Thus, the column may significantly affect the observed pulse profiles by making a dip at the phase corresponding to the position of the spot in front of the observer. 
How this would affect the results is impossible to predict without simulations and the detailed model of the accretion shock.

In our analysis we assumed a circular spot. 
This approximation is more likely valid when the magnetic inclination is small and the inner radius of accretion disc is much larger than the NS radius. 
However, for the disc close to the star and for large magnetic inclination, the spot becomes ring-like (because the matter accretes via field lines penetrating the disc within a narrow ring) or even crescent-like \citep{RU04,KR05}. 
A change in the spot shape may affect the pulse profile  \citep[see e.g.][]{PIA09,Kajava11}. 
Although, based on the simplicity of the pulse profiles, we expect this effect to be rather small.

As mentioned in Sect. {\ref{sec:methods}}, we calculate the pulse profiles using only one spot.
This is justified, because we use the observations of \source, where the accretion disc is expected to hide the second spot \citep{IP09}.
The lack of a second maximum in the pulse profile is an evidence for this. 
However, there could still be a small contribution from the second spot.
In future work and when modelling also other sources, the second spot, along with a model of light propagation to the observer through the truncated accretion disc, should be included.
 
\subsection{Additional observational constraints}\label{sec:add_obs_constr}
As mentioned in Sect. \ref{sec:vartr} and \ref{sec:datarealfit}, there are independent constraints, for instance on the observer inclination $i$ and on the interstellar hydrogen column density $N_{\mathrm{H}}$ for \source.
Using X-ray spectroscopy with \textit{XMM-Newton} \citet{PKC13} found out that $N_{\mathrm{H}} = 1.40^{+0.03}_{-0.03} \times 10^{21} ~\mathrm{cm}^{-2}$ with $1\sigma$ errors.
On the other hand, the inclination and the mass of the NS are connected and constrained via the mass function $f_{\mathrm{M}} = M\sin^{3}i/(1+q)^{2}$, where $q$ is the binary mass ratio.
Using optical observations of the companion star, during the 2008 outburst, 
\citet{ERC09} constrained the mass ratio of the system to $q=0.044^{+0.005}_{-0.004}$ and the mass function to $f_{\mathrm{M}} = 0.44^{+0.16}_{-0.13} ~\msun$.
\citet{CDM09} used the same dataset, but ended up to less realistic mass estimates. 
Later, \citet{WBH13} got similar estimates as \citet{ERC09}, using optical observations of the companion during quiescence.

We have studied how these constraints affect our results by making an additional analysis to the \source \ data with a fixed value of $N_{\mathrm{H}} = 0.14 \times 10^{22} ~\mathrm{cm}^{-2}$ and using a Gaussian 2-dimensional prior probability distribution $M\sin^{3}i = 0.48^{+0.15}_{-0.15} ~\msun$ for the mass and the inclination.
In case of the model where mass is a free parameter, using the new 2-dimensional prior we get smaller and more realistic inclinations, around 50\degr.
However, they are still not small enough to make the prior prefer larger masses (and therefore radii) of the NS.   
The smaller inclination mainly affects the spot co-latitude $\theta$, which now increased above $20 \degr$ in order to still fit the observed pulse amplitude.
Also, the relative reflection from the accretion disc $X_{\mathrm{refl}}$ becomes smaller and is now constrained below $0.05$.
When a fixed mass grid with the new priors was used, we were able to get inclinations significantly below $50 \degr$, as expected in case of higher masses.
However, the resulting posteriors for radii did not change significantly.

To get the same mass function as \citet{WBH13} or \citet{ERC09}, we would need $i$ smaller than 50\degr for a NS with a mass $1.4\msun$.
Since the observed small pulse amplitude requires large enough separation between $i$ and $\theta$, this inclination is not very easy to produce without fixing the mass (assuming that we still exclude the smallest $i$ and hence the case of almost equatorial spot).
Of course, as seen in Sect. \ref{sec:anapprox} and in Eq. \eqref{eq:amp1}, the relative pulse amplitude is also determined by the beaming parameter $h$ and the spot size $\rho$.
This might suggest that our spot size is too small, for example due to the possibly diffuse boundary of the spot, or that the beaming model needs to be improved.
In any case, the model is not perfect, since we have also neglected, for instance, absorption of light by the accretion column.

\section{Summary}
\label{sec:summary}

We have presented a method to constrain masses and radii of NSs using both spectral and X-ray timing information from AMPs. 
We have modelled the NS spectra and pulse profiles in case of two different synthetic data. 
In the first case, we assumed that the observer inclination and spot inclination angles are unknown, while in the second case we assumed that there exist constraints on both angles, for example, from the X-ray polarization measurements.  
The results showed that our method works in both cases, but the second case led to slightly tighter constraints on mass and radius.
Accuracy improved from 13\% to 8\% in determination of the mass and from 14\% to 12\% in determination of the radius (with 95\% credibilities).

We have also fitted our model to the already existing RXTE data on \source. 
The obtained constraints for the mass and the radius are in good agreement with previous studies of this object, favouring relatively small NSs.
Because the smallest masses and radii allowed by our model are not realistic for any modern equation of state, we have computed the probability distributions using also a fixed mass grid.
The results showed, that the radius can be well constrained if the NS mass is known a priori.
For example, a NS mass of $1.7~\msun$ corresponds to the equatorial radius of $\req = 11.9^{+0.5 ~(1.0)}_{-0.4 ~(1.0)}$~km.
Getting reasonable constraints for the NS radius is thus possible using even the already existing data from RXTE.

\section*{Acknowledgments}
This research was supported by the University of Turku Graduate School in Physical and Chemical Sciences (TS) and by the grant 14.W03.31.0021 of the Ministry of Education and Science of the Russian Federation (JP).
The financial support from the Magnus Ehrnrooth Foundation is also gratefully acknowledged (TS). 
We also thank Cole Miller for a useful discussion and the referee for a profound work that helped us to improve the paper. 
This research was undertaken on Finnish Grid Infrastructure (FGI) resources.


\begin{thebibliography}{63}
\expandafter\ifx\csname natexlab\endcsname\relax\def\natexlab#1{#1}\fi

\bibitem[{{AlGendy} \& {Morsink}(2014)}]{AGM14}
{AlGendy}, M. \& {Morsink}, S.~M. 2014, \apj, 791, 78
\bibitem[{{Arnaud}(1996)}]{Arn96}
{Arnaud}, K.~A. 1996, in Astronomical Data Analysis Software and Systems V, eds. G.~H. {Jacoby} \& J.~{Barnes} (San Francisco: ASP), ASP Conf. Ser., 101, 17
\bibitem[{{Baub{\"o}ck} {et~al.}(2012){Baub{\"o}ck}, {Psaltis}, {{\"O}zel}, \&  {Johannsen}}]{BPOJ12}
{Baub{\"o}ck}, M., {Psaltis}, D., {{\"O}zel}, F., \& {Johannsen}, T. 2012,  \apj, 753, 175
\bibitem[{{Beloborodov}(2002)}]{b02}
{Beloborodov}, A.~M. 2002, \apjl, 566, L85
\bibitem[{{Braje} {et~al.}(2000){Braje}, {Romani}, \& {Rauch}}]{BRR00}
{Braje}, T.~M., {Romani}, R.~W., \& {Rauch}, K.~P. 2000, \apj, 531, 447
\bibitem[{{Cackett} {et~al.}(2009){Cackett}, {Altamirano}, {Patruno}, {Miller},
  {Reynolds}, {Linares}, \& {Wijnands}}]{CAP09}
{Cackett}, E.~M., {Altamirano}, D., {Patruno}, A., {et~al.} 2009, \apjl, 694, L21
\bibitem[{{Cadeau} {et~al.}(2007){Cadeau}, {Morsink}, {Leahy}, \&  {Campbell}}]{CMLC07}
{Cadeau}, C., {Morsink}, S.~M., {Leahy}, D., \& {Campbell}, S.~S. 2007, \apj,  654, 458
\bibitem[{{Cash}(1979)}]{cash1979}
{Cash}, W. 1979, \apj, 228, 939
\bibitem[{{Chakrabarty} \& {Morgan}(1998)}]{CM98}
{Chakrabarty}, D. \& {Morgan}, E.~H. 1998, \nat, 394, 346
\bibitem[{{Chen} \& {Shaham}(1989)}]{CS89}
{Chen}, K. \& {Shaham}, J. 1989, \apj, 339, 279
\bibitem[{{Cornelisse} {et~al.}(2009){Cornelisse}, {D'Avanzo},
  {Mu{\~n}oz-Darias}, {Campana}, {Casares}, {Charles}, {Steeghs}, {Israel}, \&
  {Stella}}]{CDM09}
{Cornelisse}, R., {D'Avanzo}, P., {Mu{\~n}oz-Darias}, T., {et~al.} 2009, \aap,
  495, L1 
\bibitem[{{Deloye} {et~al.}(2008){Deloye}, {Heinke}, {Taam}, \&  {Jonker}}]{DHT08}
{Deloye}, C.~J., {Heinke}, C.~O., {Taam}, R.~E., \& {Jonker}, P.~G. 2008, \mnras, 391, 1619
\bibitem[{{Elebert} {et~al.}(2009){Elebert}, {Reynolds}, {Callanan}, {Hurley},
  {Ramsay}, {Lewis}, {Russell}, {Nord}, {Kane}, {Depoy}, \& {Hakala}}]{ERC09}
{Elebert}, P., {Reynolds}, M.~T., {Callanan}, P.~J., {et~al.} 2009, \mnras, 395, 884 
  \bibitem[{{Falanga}(2008)}]{2008AIPC.1054..157F}
{Falanga}, M. 2008, in Cool Discs, Hot Flows: The Varying Faces of Accreting Compact Objects, ed. {M.~Axelsson},  AIP Conf. Ser., 1054, 157 
\bibitem[{{Feroz} {et~al.}(2009){Feroz}, {Hobson}, \& {Bridges}}]{multinest09}
{Feroz}, F., {Hobson}, M.~P., \& {Bridges}, M. 2009, \mnras, 398, 1601 
\bibitem[{{Galloway} \& {Cumming}(2006)}]{galloway2006}
{Galloway}, D.~K. \& {Cumming}, A. 2006, \apj, 652, 559
\bibitem[{{Garc{\'{\i}}a} {et~al.}(2013){Garc{\'{\i}}a}, {Dauser}, {Reynolds}, {Kallman}, {McClintock}, {Wilms}, \& {Eikmann}}]{GDR13}
{Garc{\'{\i}}a}, J., {Dauser}, T., {Reynolds}, C.~S., {et~al.} 2013, \apj, 768,  146
\bibitem[{{Gierli{\'n}ski} {et~al.}(2002){Gierli{\'n}ski}, {Done}, \&  {Barret}}]{GDB02}
{Gierli{\'n}ski}, M., {Done}, C., \& {Barret}, D. 2002, \mnras, 331, 141
\bibitem[{{Gierli{\'n}ski} \& {Poutanen}(2005)}]{GP05}
{Gierli{\'n}ski}, M., \& {Poutanen}, J. 2005, \mnras, 359, 1261
\bibitem[{{Gilfanov} {et~al.}(1998){Gilfanov}, {Revnivtsev}, {Sunyaev}, \&  {Churazov}}]{GR98}
{Gilfanov}, M., {Revnivtsev}, M., {Sunyaev}, R., \& {Churazov}, E. 1998, \aap, 338, L83
\bibitem[{{Goodman} \& {Weare}(2010)}]{goodman10}
{Goodman}, J. \& {Weare}, J. 2010, Comm. Applied Math.  Comp. Sci., 5, 65 
\bibitem[{{Grinstead} \& {Snell}(1997)}]{GS97}
{Grinstead}, C.~M. \& {Snell}, J.~L. 1997, Introduction to Probability, 2nd ed.  (Providence, RI: American Mathematical Society)
\bibitem[{{Hartman} {et~al.}(2008){Hartman}, {Patruno}, {Chakrabarty},  {Kaplan}, {Markwardt}, {Morgan}, {Ray}, {van der Klis}, \&  {Wijnands}}]{HPC08}
{Hartman}, J.~M., {Patruno}, A., {Chakrabarty}, D., {et~al.} 2008, \apj, 675, 1468
\bibitem[{{Hou} {et~al.}(2012){Hou}, {Goodman}, {Hogg}, {Weare}, \&  {Schwab}}]{Hou12}
{Hou}, F., {Goodman}, J., {Hogg}, D.~W., {Weare}, J., \& {Schwab}, C. 2012,  \apj, 745, 198
\bibitem[{{Huijser} {et~al.}(2015){Huijser}, {Goodman}, \&  {Brewer}}]{huijser2015}
{Huijser}, D., {Goodman}, J., \& {Brewer}, B.~J. 2015, arXiv:1509.02230
\bibitem[{{Ibragimov} \& {Poutanen}(2009)}]{IP09}
{Ibragimov}, A. \& {Poutanen}, J. 2009, \mnras, 400, 492
\bibitem[{{Jahoda} {et~al.}(2006){Jahoda}, {Markwardt}, {Radeva}, {Rots},
  {Stark}, {Swank}, {Strohmayer}, \& {Zhang}}]{JMR06}
{Jahoda}, K., {Markwardt}, C.~B., {Radeva}, Y., {et~al.} 2006, \apjs, 163, 401
\bibitem[{{Kajava} {et~al.}(2011){Kajava}, {Ibragimov}, {Annala}, {Patruno}, \&  {Poutanen}}]{Kajava11}
{Kajava}, J.~J.~E., {Ibragimov}, A., {Annala}, M., {Patruno}, A., \&  {Poutanen}, J. 2011, \mnras, 417, 1454
\bibitem[{{Kolehmainen} {et~al.}(2011){Kolehmainen}, {Done}, \& {D{\'{\i}}az  Trigo}}]{Kolehmainen11}
{Kolehmainen}, M., {Done}, C., \& {D{\'{\i}}az Trigo}, M. 2011, \mnras, 416, 311
  \bibitem[{{Kulkarni} \& {Romanova}(2005)}]{KR05}
{Kulkarni}, A.~K. \& {Romanova}, M.~M. 2005, \apj, 633, 349
\bibitem[{{Lattimer}(2012)}]{Lattimer12ARNPS}
{Lattimer}, J.~M. 2012, ARNPS, 62, 485
\bibitem[{{Lattimer} \& {Prakash}(2004)}]{LP04}
{Lattimer}, J.~M. \& {Prakash}, M. 2004, Sci, 304, 536
\bibitem[{{Leahy} {et~al.}(2008){Leahy}, {Morsink}, \& {Cadeau}}]{LMC08}
{Leahy}, D.~A., {Morsink}, S.~M., \& {Cadeau}, C. 2008, \apj, 672, 1119
\bibitem[{{Lindblom}(1992)}]{lindblom1992}
{Lindblom}, L. 1992, \apj, 398, 569
\bibitem[{{Lo} {et~al.}(2013){Lo}, {Miller}, {Bhattacharyya}, \&  {Lamb}}]{lomiller13}
{Lo}, K.~H., {Miller}, M.~C., {Bhattacharyya}, S., \& {Lamb}, F.~K. 2013, \apj, 776, 19
\bibitem[{{Lo} {et~al.}(2018){Lo}, {Miller}, {Bhattacharyya}, \&  {Lamb}}]{lo_err18}
{Lo}, K.~H., {Miller}, M.~C., {Bhattacharyya}, S., \& {Lamb}, F.~K. 2018, \apj,  854, 187
\bibitem[{{Magdziarz} \& {Zdziarski}(1995)}]{MZ95}
{Magdziarz}, P. \& {Zdziarski}, A.~A. 1995, \mnras, 273, 837
\bibitem[{{Miller} \& {Lamb}(1998)}]{ML98}
{Miller}, M.~C. \& {Lamb}, F.~K. 1998, \apjl, 499, L37
\bibitem[{{Miller} \& {Lamb}(2015)}]{ML15}
{Miller}, M.~C. \& {Lamb}, F.~K. 2015, \apj, 808, 31
\bibitem[{{Misner} {et~al.}(1973){Misner}, {Thorne}, \& {Wheeler}}]{mtw73}
{Misner}, C.~W., {Thorne}, K.~S., \& {Wheeler}, J.~A. 1973, {Gravitation} (San  Francisco: W.H.~Freeman and Co.)
\bibitem[{{Morsink} \& {Leahy}(2011)}]{ML11}
{Morsink}, S.~M. \& {Leahy}, D.~A. 2011, \apj, 726, 56
\bibitem[{{Morsink} {et~al.}(2007){Morsink}, {Leahy}, {Cadeau}, \&  {Braga}}]{MLC07}
{Morsink}, S.~M., {Leahy}, D.~A., {Cadeau}, C., \& {Braga}, J. 2007, \apj, 663,  1244
\bibitem[{{N{\"a}ttil{\"a}} \& {Pihajoki}(2018)}]{NP18}
{N{\"a}ttil{\"a}}, J. \& {Pihajoki}, P. 2018, \aap, in press (arxiv:1709.07292)
\bibitem[{{{\"O}zel} {et~al.}(2016){{\"O}zel}, {Psaltis}, {Arzoumanian}, {Morsink}, \& {Baub{\"o}ck}}]{OPA16}
{{\"O}zel}, F., {Psaltis}, D., {Arzoumanian}, Z., {Morsink}, S., \& {Baub{\"o}ck}, M. 2016, \apj, 832, 92
\bibitem[{{Page}(1995)}]{Page95}
{Page}, D. 1995, \apj, 442, 273
\bibitem[{{Pechenick} {et~al.}(1983){Pechenick}, {Ftaclas}, \& {Cohen}}]{PFC83}
{Pechenick}, K.~R., {Ftaclas}, C., \& {Cohen}, J.~M. 1983, \apj, 274, 846
\bibitem[{{Pinto} {et~al.}(2013){Pinto}, {Kaastra}, {Costantini}, \& {de
  Vries}}]{PKC13}
{Pinto}, C., {Kaastra}, J.~S., {Costantini}, E., \& {de Vries}, C. 2013, \aap,  551, A25
\bibitem[{{Poutanen}(2008)}]{P08}
{Poutanen}, J. 2008, in A Decade of Accreting Millisecond X-ray Pulsars, eds. R.~{Wijnands},  D.~{Altamirano}, P.~{Soleri}, N.~{Degenaar}, N.~{Rea}, P.~{Casella},
  A.~{Patruno}, \& M.~{Linares}, AIP Conf. Ser., 1068, 77 
\bibitem[{{Poutanen} \& {Beloborodov}(2006)}]{PB06}
{Poutanen}, J. \& {Beloborodov}, A.~M. 2006, \mnras, 373, 836
\bibitem[{{Poutanen} \& {Gierli{\'n}ski}(2003)}]{PG03}
{Poutanen}, J. \& {Gierli{\'n}ski}, M. 2003, \mnras, 343, 1301
\bibitem[{{Poutanen} {et~al.}(2009){Poutanen}, {Ibragimov}, \& {Annala}}]{PIA09}
{Poutanen}, J., {Ibragimov}, A., \& {Annala}, M. 2009, \apjl, 706, L129
\bibitem[{{Psaltis} {et~al.}(2014){Psaltis}, {{\"O}zel}, \& {Chakrabarty}}]{POC14}
{Psaltis}, D., {{\"O}zel}, F., \& {Chakrabarty}, D. 2014, \apj, 787, 136 
\bibitem[{{Romanova} {et~al.}(2004){Romanova}, {Ustyugova}, {Koldoba}, \&  {Lovelace}}]{RU04}
{Romanova}, M.~M., {Ustyugova}, G.~V., {Koldoba}, A.~V., \& {Lovelace}, R.~V.~E. 2004, \apj, 610, 920
\bibitem[{{Steiner} {et~al.}(2009){Steiner}, {Narayan}, {McClintock}, \& {Ebisawa}}]{Steiner2009}
{Steiner}, J.~F., {Narayan}, R., {McClintock}, J.~E., \& {Ebisawa}, K. 2009,  \pasp, 121, 1279
\bibitem[{{Stevens} {et~al.}(2016){Stevens}, {Fiege}, {Leahy}, \&  {Morsink}}]{Stevens2016}
{Stevens}, A.~L., {Fiege}, J.~D., {Leahy}, D.~A., \& {Morsink}, S.~M. 2016, \apj, 833, 244
\bibitem[{{Sunyaev} \& {Titarchuk}(1980)}]{ST80}
{Sunyaev}, R.~A., \& {Titarchuk}, L.~G. 1980, \aap, 86, 121
\bibitem[{{Viironen} \& {Poutanen}(2004)}]{VP04}
{Viironen}, K. \& {Poutanen}, J. 2004, \aap, 426, 985
\bibitem[{{Wang} {et~al.}(2013){Wang}, {Breton}, {Heinke}, {Deloye}, \&
  {Zhong}}]{WBH13}
{Wang}, Z., {Breton}, R.~P., {Heinke}, C.~O., {Deloye}, C.~J., \& {Zhong}, J.
  2013, \apj, 765, 151
\bibitem[{{Watts} {et~al.}(2016){Watts}, {Andersson}, {Chakrabarty}, {Feroci}, {Hebeler}, {Israel}, {Lamb}, {Miller}, {Morsink}, {{\"O}zel}, {Patruno},  {Poutanen}, {Psaltis}, {Schwenk}, {Steiner}, {Stella}, {Tolos}, \& {van der Klis}}]{2016RvMP...88b1001W}
{Watts}, A.~L., {Andersson}, N., {Chakrabarty}, D., {et~al.} 2016, Rev. Mod. Phys., 88, 021001
\bibitem[{{Weinberg} {et~al.}(2001){Weinberg}, {Miller}, \&  {Lamb}}]{WM01}
{Weinberg}, N., {Miller}, M.~C., \& {Lamb}, D.~Q. 2001, \apj, 546, 1098
\bibitem[{{Weisskopf} {et~al.}(2016){Weisskopf}, {Ramsey}, {O'Dell}, {Tennant},
  {Elsner}, {Soffitta}, {Bellazzini}, {Costa}, {Kolodziejczak}, {Kaspi},
  {Muleri}, {Marshall}, {Matt}, \& {Romani}}]{IXPE}
{Weisskopf}, M.~C., {Ramsey}, B., {O'Dell}, S., {et~al.} 2016, in \procspie,
  Vol. 9905, Space Telescopes and Instrumentation 2016: Ultraviolet to Gamma Ray, 990517
\bibitem[{{Wijnands} \& {van der Klis}(1998)}]{WvdK98}
{Wijnands}, R. \& {van der Klis}, M. 1998, \nat, 394, 344
\bibitem[{{Zhang} {et~al.}(2016){Zhang}, {Feroci}, {Santangelo}, {Dong},
  {Feng}, {Lu}, {Nandra}, {Wang}, {Zhang}, {Bozzo}, {Brandt}, {De Rosa}, {Gou},
  {Hernanz}, {van der Klis}, {Li}, {Liu}, {Orleanski}, {Pareschi}, {Pohl},
  {Poutanen}, {Qu}, {Schanne}, {Stella}, {Uttley}, {Watts}, {Xu}, {Yu}, {in 't
  Zand}, {Zane}, {Alvarez}, {Amati}, {Baldini}, {Bambi}, {Basso},
  {Bhattacharyya S.}, {}, {Belloni}, {Bellutti}, {Bianchi}, {Brez}, {Bursa},
  {Burwitz}, {Budtz-J{\o}rgensen}, {Caiazzo}, {Campana}, {Cao}, {Casella},
  {Chen}, {Chen}, {Chen}, {Chen}, {Chen}, {Chen}, {Civitani}, {Coti Zelati},
  {Cui}, {Cui}, {Dai}, {Del Monte}, {de Martino}, {Di Cosimo}, {Diebold},
  {Dovciak}, {Donnarumma}, {Doroshenko}, {Esposito}, {Evangelista}, {Favre},
  {Friedrich}, {Fuschino}, {Galvez}, {Gao}, {Ge}, {Gevin}, {Goetz}, {Han},
  {Heyl}, {Horak}, {Hu}, {Huang}, {Huang}, {Hudec}, {Huppenkothen}, {Israel},
  {Ingram}, {Karas}, {Karelin}, {Jenke}, {Ji}, {Korpela}, {Kunneriath},
  {Labanti}, {Li}, {Li}, {Li}, {Liang}, {Limousin}, {Lin}, {Ling}, {Liu},
  {Liu}, {Liu}, {Lu}, {Lund}, {Lai}, {Luo}, {Luo}, {Ma}, {Mahmoodifar},
  {Marisaldi}, {Martindale}, {Meidinger}, {Men}, {Michalska}, {Mignani},
  {Minuti}, {Motta}, {Muleri}, {Neilsen}, {Orlandini}, {Pan}, {Patruno},
  {Perinati}, {Picciotto}, {Piemonte}, {Pinchera}, {Rachevski A.}, {Rapisarda},
  {Rea}, {Rossi}, {Rubini}, {Sala}, {Shu}, {Sgro}, {Shen}, {Soffitta}, {Song},
  {Spandre}, {Stratta}, {Strohmayer}, {Sun}, {Svoboda}, {Tagliaferri},
  {Tenzer}, {Hong}, {Taverna}, {Torok}, {Turolla}, {Vacchi}, {Wang}, {Walton},
  {Wang}, {Wang}, {Wang}, {Wang}, {Weng}, {Wilms}, {Winter}, {Wu}, {Wu},
  {Xiong}, {Xu}, {Xue}, {Yan}, {Yang}, {Yang}, {Yang}, {Yuan}, {Yuan}, {Yuan},
  {Zampa}, {Zampa}, {Zdziarski}, {Zhang}, {Zhang}, {Zhang}, {Zhang}, {Zhang},
  {Zhang}, {Zheng}, {Zhou}, \& {Zhou X.~L.}}]{EXTP}
{Zhang}, S.~N., {Feroci}, M., {Santangelo}, A., {et~al.} 2016, in \procspie,
  Vol. 9905, Space Telescopes and Instrumentation 2016: Ultraviolet to Gamma Ray, 99051Q
\end{thebibliography}

\begin{appendix}
\section{Numerical solutions for light bending and time delay integrals}
\label{sec:appendix}
 
Numerical calculations using directly integral (\ref{eq:bend2}) are not possible. 
Making substitution $y=R/r$, the integral becomes 
\begin{equation}\label{eq:bend3} 
  \psi_{\rm p} (R,\alpha)= {\sin\alpha} \ \int_0^{1} d y \ 
  \left[ 1- u  - y^2 \sin^2\!\alpha\ (1-yu)\right]^{-1/2}.
\end{equation}
Applying quadrature formulae to this form of the integral leads to dramatic errors when $\sin\alpha\sim 1$, because of the divergent behaviour of the integrand at $y=1$. 
However, simple variable substitution $x=\sqrt{1-y}$ removes the divergence: 
\begin{equation}\label{eq:bend_nodiver}
  \psi_{\rm p} (R,\alpha)=2\frac{\sin\alpha}{\sqrt{1-u}}\ \int_0^{1} 
 \left( \cos^2\!\alpha + x^2 q \right)^{-1/2} \, x\ d x , 
\end{equation}
where 
 \begin{equation}\label{eq:q_par_def}
q= \left[ 2-x^2 - u (1-x^2)^2/(1-u)  \right]   \sin^2\!\alpha .
\end{equation}
Notice that now for $\cos\alpha=0$, the integrand does not diverge at $x=0$ and the integral can be computed without problems using, for example, Simpson quadratures.

Also, direct calculations using expression (\ref{eq:timedelay}) are not possible. 
Making substitution $y=R/r$, the integral becomes 
\begin{equation}\label{eq:delay_1}
\Delta t_{\rm p} = \frac{R}{c} \int_0^{1} \frac{d y}{y^2(1-yu)} 
\left\{ \left[ 1- \sin^2\!\alpha\ y^2 (1-yu)/(1-u) \right] ^{-1/2}  -1 \right\} .
\end{equation}
This form  has seemingly two divergences at $y=0$ and for $\sin\alpha=1$ at $y=1$. 
The first divergence is removed by multiplying and dividing by the expression similar to that in the curly brackets but with the $+1$ in the end. 
This gives
 \begin{equation}\label{eq:delay_nodiver_yzero}
\Delta t_{\rm p} = \frac{R}{c}\ \frac{\sin^2\!\alpha}{1-u} 
                             \int_0^{1} \frac{\ d y}{\sqrt{1 - w} 
                             \left( 1+\sqrt{1-w } \right)} ,
\end{equation}
where
 \begin{equation}\label{eq:w_par_def}
w= \frac{ y^2(1-yu)}{1-u} \sin^2\alpha .
\end{equation}
The divergence at $y=1$ (for $\sin\alpha = 1 $) is of the $1/\sqrt{1-y}$\ -type and therefore is trivially 
removed by variable substitution $x=\sqrt{1-y}$.  
This transforms the integral to 
 \begin{equation}\label{eq:delay_nodiver}
\Delta t_{\rm p} = 2 \frac{R}{c}\ \frac{\sin^2\!\alpha}{1-u} 
                             \int_0^{1} \frac{x\ d x}{\sqrt{x^2  q + \cos^2\!\alpha} 
                             \left( 1+\sqrt{x^2 q + \cos^2\!\alpha } \right)} ,
\end{equation}
where $q$ is given by Eq. \eqref{eq:q_par_def}. 
Now for $\cos\alpha=0$, the integrand does not diverge at all, but becomes $1/(qx+\sqrt{q})$ 
and the integral can be computed using e.g. Simpson quadratures.

\end{appendix}

\end{document}